\documentclass[sigconf, nonacm]{acmart}

\AtBeginDocument{%
  \providecommand\BibTeX{{%
    \normalfont B\kern-0.5em{\scshape i\kern-0.25em b}\kern-0.8em\TeX}}}

\usepackage{tabularx}
\usepackage{multirow}
\usepackage{color}
\usepackage{listings}
\newcommand{\ie}{i.\,e.}
\newcommand{\eg}{e.\,g.}

\newcommand\changed[1]{{{#1}}}

\newcommand{\percent}{\%}
\newcommand*{\num}[1]{\ensuremath{#1}}
\newcommand*{\SI}[2]{\ensuremath{#1\,#2}}

\begin{document}


\title[Privacy Implications and User Acceptance of COVID-19-Related Smartphone Apps]{Apps Against the Spread: Privacy Implications and User Acceptance of COVID-19-Related Smartphone Apps on Three Continents}


\author{Christine Utz}
\orcid{0000-0003-4346-6911}
\affiliation{%
    \institution{Ruhr University Bochum}
    \city{Bochum}
    \country{Germany}
}
\email{christine.utz@rub.de}

\author{Steffen Becker}
\orcid{0000-0001-7526-5597}
\affiliation{%
    \institution{Ruhr University Bochum}
    \city{Bochum}
    \country{Germany}
}
\affiliation{%
\institution{Max Planck Institute for Security and Privacy}
    \city{Bochum}
    \country{Germany}
}
\email{steffen.becker@rub.de}

\author{Theodor Schnitzler}
\orcid{0000-0001-7575-1229}
\affiliation{%
    \institution{Ruhr University Bochum}
    \city{Bochum}
    \country{Germany}
}
\email{theodor.schnitzler@rub.de}

\author{Florian M. Farke}
\orcid{0000-0001-7138-4978}
\affiliation{%
    \institution{Ruhr University Bochum}
    \city{Bochum}
    \country{Germany}
}
\email{florian.farke@rub.de}

\author{Franziska Herbert}
\orcid{}
\affiliation{%
    \institution{Ruhr University Bochum}
    \city{Bochum}
    \country{Germany}
}
\email{franziska.herbert@rub.de}

\author{Leonie Schaewitz}
\orcid{0000-0002-4339-501X}
\affiliation{%
    \institution{Ruhr University Bochum}
    \city{Bochum}
    \country{Germany}
}
\email{leonie.schaewitz@rub.de}

\author{Martin Degeling}
\orcid{0000-0001-7048-781X}
\affiliation{%
    \institution{Ruhr University Bochum}
    \city{Bochum}
    \country{Germany}
}
\email{martin.degeling@rub.de}

\author{Markus D\"{u}rmuth}
\orcid{}
\affiliation{%
    \institution{Ruhr University Bochum}
    \city{Bochum}
    \country{Germany}
}
\email{markus.duermuth@rub.de}

\renewcommand{\shortauthors}{Utz, Becker, Schnitzler, Farke, Herbert, Schaewitz, Degeling, and D\"{u}rmuth}

\begin{abstract}
The COVID-19 pandemic has fueled the development of smartphone applications to assist disease management. Many ``corona apps'' require widespread adoption to be effective, which has sparked public debates about the privacy, security, and societal implications of government-backed health applications.
We conducted a representative online study in Germany (n = 1003), the US (n = 1003), and China (n = 1019) to investigate user acceptance of corona apps, using a vignette design based on the contextual integrity framework. We explored apps for contact tracing, symptom checks, quarantine enforcement, health certificates, and mere information.
Our results provide insights into data processing practices that foster adoption and reveal significant differences between countries, with user acceptance being highest in China and lowest in the US. Chinese participants prefer the collection of personalized data, while German and US participants favor anonymity. Across countries, contact tracing is viewed more positively than quarantine enforcement, and technical malfunctions negatively impact user acceptance.
\end{abstract}

\ccsdesc[500]{Security and privacy~Social aspects of security and privacy}
\ccsdesc[300]{Security and privacy~Domain-specific security and privacy architectures}
\ccsdesc[300]{Human-centered computing~Empirical studies in HCI}

\keywords{digital contact tracing, {COVID-19}, privacy}

\maketitle

\section{Introduction}
\label{sec:intro}

In early 2020, the global pandemic of coronavirus disease 2019 (COVID-19), caused by severe acute respiratory syndrome coronavirus 2 (SARS‑CoV-2), has driven the development of digital tools to assist traditional disease management methods in limiting the spread of the novel coronavirus and future pathogens. Since then, numerous software projects, especially involving smartphone applications, have been launched by public and private entities around the world, with specific purposes that greatly vary, but all with the common goal to help contain the spread of the pandemic.

These include smartphone apps for \emph{digital contact tracing}, which was identified to be an effective method to assist health authorities in breaking chains of infection~\cite{ferretti_contacttracing_2020}. It uses proximity information from smartphones to determine which other users of compatible systems a person has been in recent close contact with, and if a user tests positive for coronavirus, these contacts can be notified to encourage timely isolation and testing. Among the first smartphone apps released for this purpose was Singapore's TraceTogether app, launched on March 20, 2020~\cite{sggov_tracetogether_2020}, and multiple countries around the globe have since followed suit.
Other regions developed digital \emph{health certificate} systems, first introduced in February 2020 in the eastern Chinese city of Hangzhou ~\cite{mozur_china_2020}. Checkpoints could require people to present their personal QR code displayed in the associated app and only let them pass if the system considered them low-risk. The travel industry has expressed an interest in similar systems to reestablish international travel for people with a recent negative test result~\cite{meneguzzi_immunity_2020}.
Hong Kong issued a smartphone app, combined with a QR code wristband, to ensure that people do not break the government-mandated \emph{quarantine} for new arrivals~\cite{hkgov_stayhomesafe_2020}.
Other countries including South Korea and Australia issued \emph{symptom check} apps that allow users to monitor their health for possible symptoms of COVID-19~\cite{augov_coronavirusaustralia_2020, mohw_selfcheck_2020}. The Australian app also serves as an example for countless apps issued by both public and private organizations to provide \emph{information} about the virus, its spread, rules of hygiene, and local regulations.

The use of smartphone applications in the fight against the pandemic has sparked intense public debates about the privacy, security, and societal implications of government-recommended health apps and specific aspects of their implementations.
In Europe, the debate mainly focused on the selection of a centralized or a decentralized architecture for digital contact tracing~\cite{criddle_split_2020}, which involves questions of \emph{data transmission} and the \emph{data recipient} and ultimately led to multiple countries adopting a decentralized approach due to privacy concerns. 
The Chinese health code system has raised questions about the consequences of \emph{technical} issues and long-term \emph{societal implications} of apps used to regulate movement for the sake of disease prevention~\cite{kuo_monitoring_2020}. The quarantine enforcement app issued by the Polish government was subject to criticism due to the \emph{collected data}, technical malfunctions, and \emph{data retention} of up to six years~\cite{scott_poland_2020}. 

The public interest in the organizational and technical details of these ``corona apps'' is no surprise: Since their efficiency may depend on widespread voluntary adoption~\cite{ferretti_contacttracing_2020}, many were recommended for general adoption by governments and health authorities. User acceptance and their willingness to use these apps can thus be critical for the overall success of smartphone apps in the efforts to fight the pandemic. 
Many of the publicly discussed and criticized aspects are well-known elements in the theory of ``privacy as contextual integrity'', which acknowledges that factors beyond the technical implementation, like social norms and expectations, influence perceptions of privacy violations~\cite{nissenbaum_privacy_2004}.

\changed{
People's willingness to share personal data from or with their personal devices to benefit their healthcare has previously been studied in a wide range of different contexts, including mental health~\cite{dimatteo_mentalhealthapps_2018, nicholas_mentalhealth_2019}, HIV~\cite{warner_hiv_2019}, recovery from surgery~\cite{abelson_surgery_2017} and rheumatic diseases~\cite{najm_rheumatic_2019}, but these studies are typically confined to specific locations or populations.
The COVID-19 pandemic offers the unique opportunity to study on a large scale, in the context of a virus that could infect anyone at any given time, which individual and privacy-related factors influence people's willingness to use an app that not only serves individual goals but also has a societal purpose. Research in this area thus not only informs the debate around corona apps but can aid the design of health-related apps that would benefit from widespread adoption outside the specific context of the pandemic. 
As early as spring 2020, when the first smartphone apps for digital contact tracing were rolled out, the HCI community has taken this opportunity to investigate people's willingness to use mobile apps to help fight the COVID-19 pandemic, focusing on apps for digital contact tracing~\cite{zhang_covid_privacy_2020, altmann_acceptability_2020, simko_contact_tracing_2020, trang_oneapp_2020, li_covid_apps_2020, kaptchuk_covid19apps_2020}. With countries around the world also having issued apps for different purposes in the pandemic, it is interesting to study factors for the adoption of corona apps for purposes beyond contact tracing and how these factors differ between app types.
}
Public debates can lead to different outcomes in different countries, influenced by cultural differences, trust in technology and institutions, and regional developments of the pandemic. It is therefore insightful to study user acceptance of corona apps in different regions.

In this work, we present the results of an online survey conducted in Germany, the US, and China between June and August 2020 to investigate user acceptance of corona apps and the factors that influence people's willingness to use them. These countries were chosen because they differ with respect to their perception of privacy and the state of the COVID-19 pandemic.
Our study is based on a vignette design of hypothetical corona apps and inspired by existing apps from around the world and the contextual integrity framework.

Our research makes the following contributions to the HCI community:
\begin{itemize}
    \setlength{\itemsep}{0pt}
    \setlength{\parskip}{0pt}
    \setlength{\parsep}{0pt}
    \item We expand existing knowledge about user perception and acceptance of COVID-19-related smartphone apps beyond the purpose of digital contact tracing and investigate people's willingness to use them for other purposes found in government-issued corona apps, \ie, quarantine enforcement, symptom checks, health certificates, and information. Our data shows that contact tracing apps have the highest public support across all countries.
    \item We extend existing research on the acceptability of corona apps beyond the Western world and conduct the study in China, where the pandemic first started and the use of smartphone apps for health screening and movement control is widespread. While in all surveyed countries the most widely discussed and used apps see more support, acceptance in China is higher and not related to individual concerns or the expected benefits.
    \item Our results show that there is skepticism towards apps provided by the government that results in 15 to 21\,\% of participants not willing to use any app in Germany and the US, while in China users are more concerned that apps might raise stress levels through continued awareness.
\end{itemize}


\section{Contextual Integrity}

Previous studies have found that privacy is often not the main concern of users when deciding whether to adopt a certain technology or not. While this behavior is frequently attributed to the privacy paradox~\cite{norberg_privacyparadox_2007, gerber_privacy_2018} -- participants express high privacy concern but rarely act accordingly -- researchers increasingly argue that the trade-off is more complex. According to Nissenbaum's theory of privacy as contextual integrity~(CI), individual privacy preferences and decisions can be described with respect to the appropriateness of information flows~\cite{nissenbaum_privacy_2004}. What constitutes an appropriate flow is determined by different factors, such as actors (Whose data is involved? Who will send and receive it?), information types (What types of data are concerned?), and transmission principles (What are the technical and organizational means of the data transmission?). The examples of public debates in Section~\ref{sec:intro}, such as the Polish quarantine app or the discussion about a (de)centralized infrastructure, illustrate that public debates about the privacy implications of new technology often implicitly refer to these factors. 

\changed{
The understanding of privacy as contextual integrity is based on philosophical privacy theories that go beyond the famous ``right to be let alone''~\cite{warren_privacy_1890} and frames privacy not as an individual right with fixed boundaries (\eg, the private home vs. the public square) but as a concept that focuses on the role of information in relations. This focus on informational privacy is useful in the context of the digital technologies we consider in this study, but it is worth noting that other dimensions of privacy like decisional and locational privacy~\cite{roessler_privacy_2006} as well as physical, social, and psychological privacy~\cite{burgoon_privacy_1982} are discussed in the literature.
The idea of ``appropriateness'' in CI highlights that privacy is a contested problem~\cite{mulligan_contested_2016} and that our understanding of privacy and norms associated with it are continuously changing. 
Because of this, privacy as contextual integrity has seen wide adoption in computer science~\cite{benthall_contextual_2017} and HCI research~\cite{barkhuus_mismeasurement_2012} and has also been used in the context of health data~\cite{patterson_healthinformationflows_2013, nicholas_mentalhealth_2019}. Moreover, it was recently suggested as an appropriate research framework to evaluate potential long-term risks of {COVID-19}-related surveillance technologies~\cite{vitak_covid19surveillance_2020}.
For this research, the contextual integrity framework allows us to operationalize factors influencing privacy perceptions and norms so that we are able to construct relevant app scenarios for our study (see Section~\ref{sec:existing-apps}). 
Our study then shows whether this operationalization in factors can actually help to better understand the importance of each factor for privacy decision making.
}

\section{Related Work}

\subsection{User Perceptions of Digital Contact Tracing Apps in the COVID-19 Pandemic}

Since the emergence of the COVID-19 pandemic in early 2020, multiple research groups have investigated people's privacy perceptions of smartphone apps for digital contact tracing. 

One of the earliest studies was conducted by Zhang et al.~\cite{zhang_covid_privacy_2020}, who explored US Americans' perceptions of privacy and surveillance in the COVID-19 pandemic in an online survey with \num{2612} participants, administered between March 30 and April 1, 2020.
Their preliminary results indicate that Americans favor traditional contact tracing and health screenings over app-based digital contact tracing.
In a conjoint analysis the researchers identified two attributes of contact tracing apps with statistically significant effects on the reported likelihood of downloading the app: participants preferred decentralized data storage and Bluetooth proximity technology over location tracking.
Between late March and early April, 2020, Altmann et al.~\cite{altmann_acceptability_2020} conducted several representative online surveys in France, Germany, Italy, the UK, and the US to study the acceptance of app-based contact tracing. When shown the description of a hypothetical contact tracing app, \num{75}\,\% of participants indicated their willingness to install the app, with a significantly lower acceptance in Germany and the US than in the other three countries. The study found lower acceptance for participants who have less trust in their national government, and the main reasons against app installation were identified to be fear of government surveillance after the pandemic and the phone being hacked.
In early April, Simko et al.~\cite{simko_contact_tracing_2020} started a longitudinal study on contact tracing and privacy with a sequence of online surveys, each with around \num{100} participants from around the world. Their preliminary results indicate that \num{72}\,\% of participants would be at least somewhat likely to download a contact tracing app if it provides ``perfect data protection''.
In April 2020, Trang et al.~\cite{trang_oneapp_2020} conducted a survey with \num{518} participants in Germany to examine app specifications for mass acceptance of contact tracing apps in three dimensions: benefit appeal, privacy design, and convenience design. Participants were categorized into three groups based on their willingness to install a contact tracing app: critics, undecided, and advocates. For critics, self-benefit and high privacy were most important, while high convenience design had the strongest effect for undecided participants.
For advocates, all three dimensions played a subordinate role.
Between late April and early May, 2020, Li et al.~\cite{li_covid_apps_2020} surveyed \num{208} US citizens on privacy-utility trade-offs on the basis of six contact tracing app scenarios in two dimensions: centralized vs. decentralized architecture and location collection of infected users (in public places) vs. no location collection.
Their preliminary findings show that a majority of participants preferred to install apps that use centralized servers and share diagnosed users’ recent locations in public places to reveal hot spots of infection.
Kaptchuk et al.~\cite{kaptchuk_covid19apps_2020} conducted a survey with \num{789} US participants to study the influence of accuracy and privacy on the intention to install contact tracing apps.
They found that \num{70}--\num{80}\,\% of participants were willing to install an app that is perfectly private, accurate, or both, and that false negatives had a significantly stronger influence than false positives or privacy risks.

\changed{
\subsection{Sharing of Health Information}
A related, broad field of research are users' privacy perceptions and sharing behavior of health data~\cite{baig_dnatesting_2020, petelka_bipolar_2020, peng_healthapps_2016, nicholas_mentalhealth_2019, dimatteo_mentalhealthapps_2018, gorm_workplace_2016}. 
A study investigating individuals' views of sharing sensor data~\cite{nicholas_mentalhealth_2019} revealed that users were more comfortable sharing activity data such as sleep, mood, and physical activity as opposed to communication logs, location, and social activity. Users were more comfortable with sharing the mentioned data with their doctors than with family members or electronic health record systems~\cite{nicholas_mentalhealth_2019}. Patients of a clinic for mood and anxiety disorders were asked about a hypothetical app to better diagnose and treat their health disorder~\cite{dimatteo_mentalhealthapps_2018}. Most of the patients stated to be willing to install such an app. Patients were reluctant to share personal communication data (audio and SMS) but more willing to share less personal data, such as when their phone screen was turned on and off. 

Relevant factors for non-disclosure of HIV health information in the context of sex-social apps were stigmatization and the fear for discrimination and disadvantages~\cite{petelka_bipolar_2020}. People using at-home DNA testing feared surveillance by the government, saw the misuse of a third party as potential risk of sharing health DNA data and expressed the desire for transparency and some level of control over their health DNA data. Participants also showed resignation towards privacy as they felt a lack of control over their data~\cite{baig_dnatesting_2020}. 
In an interview study on the perception and use of mobile health apps, ``lack of need,'' ``lack of app literacy'' and ``lack of awareness'' were mentioned as barriers for the usage of health apps~\cite{ peng_healthapps_2016}. Users reported aspects such as ``lack of time and effort'' and too much storage space or high battery consumption as barriers to a continued usage of health apps~\cite{ peng_healthapps_2016}. 

In the context of a workplace health promotion campaign involving the use of step-counting technology, participants had only little concerns to share their activity data with organizational entities, such as their employer or technology companies, but were more concerned regarding disclosure to other individuals~\cite{gorm_workplace_2016}.The positive rhetoric of the campaign also had a positive influence on the willingness to share activity data. 
}

\subsection{User Acceptance of Information Technology}

The most common and widely applied theoretical frameworks for investigating factors that can explain usage intention and acceptance of information technology include the theory of planned behavior~\cite{ajzen_tbp_1991} and the technology acceptance model (TAM~\cite{davis_informationtechnology_1989}); Venkatesh et al.~\cite{venkatesh_acceptance_2003} provide an overview.

Factors that have been identified to directly or indirectly influence the intention to use and accept technologies (\eg, apps, social network sites, blogs) are perceived usefulness and perceived ease of use~\cite{davis_informationtechnology_1989, beldad_technologyacceptance_2017}, social influence~\cite{venkatesh_acceptance_2003, hsu_blogusage_2008}, facilitating conditions, demographic factors like gender, age and experience~\cite{venkatesh_acceptance_2003}, and trusting the party behind the technologies~\cite{beldad_location_2015}.

Another investigated factor is privacy.
Gu et al.~\cite{gu_privacy_2017} studied privacy concern for mobile applications in relation to the permissions they request. They found that an app's popularity had a positive and the overall privacy concern a negative effect on download intention. Following this research, Wottrich et al.~\cite{wottrich_privacy_2018} studied the trade-off between privacy and the value an app has for the user. They found that perceived intrusiveness and privacy concern have a negative impact on the decision whether or not to install an app, though the positive effect of the perceived value of the app is more important in the privacy calculus.
Focusing on health-related apps, Zhou et al.~ \cite{zhou_barriers_2019} surveyed and interviewed 117 smartphone users, the majority of whom had already used mobile health apps. The study found that privacy and security issues can be barriers to adoption regardless of demographic factors.
\section{Method}

To understand which data processing factors influence people's willingness to use corona apps, we conducted an online survey in Germany, the USA, and China between June and August 2020.
At the core of our study, we presented participants with ten scenarios in the form of vignettes, each describing a hypothetical corona app, and asked them to assess each app according to a set of criteria. In the following, we describe our vignette design and the rest of our questionnaire, the data collection process including the corresponding state of the pandemic in each country, and the methods used for data analysis. 

\subsection{Vignette Design}

Vignettes are ``short stories about hypothetical characters in specified circumstances, to whose situation the interviewee is invited to respond''~\cite{finch_vignette_1987} and have been used in previous work studying contextual privacy~\cite{martin_records_2017, emaminaeini_iot_2017, apthorpe_iottoys_2019}.
The vignettes at the heart of our study are composed of different \emph{factors} of data processing practices, inspired by the contextual integrity framework.
Each factor can be assigned one of multiple \emph{factor levels} we determined by examining existing corona apps.

\subsubsection{Analyzing Existing Corona Apps for Data Processing Factors}
\label{sec:existing-apps}

To ensure our study is firmly rooted in existing corona apps from around the world, three researchers investigated the space of apps and their data processing practices by systematically examining \num{43} apps from Wikipedia's list of government-endorsed corona apps as of April 27, 2020\footnote{\url{https://en.wikipedia.org/w/index.php?title=COVID-19\_apps\&oldid=953499862\#List\_of\_apps\_by\_country}.}.
For each app, we analyzed publicly available resources: listings in app stores, the app’s official website, terms of service, privacy policies, and official press announcements. 
\changed{
The retrieved information focused on the concept of data flows in the contextual integrity (CI) theory: actors (primary and secondary data recipients), information types (collected data), and transmission principles (what triggered a data transmission process, retention).
We also noted potential technical and societal implications of using an app as they were discussed in app reviews and the public debate and could influence people's willingness to install and use an app.
For systematization and documentation purposes we collected the following app metadata: name and country of the app, its development process and status, and the underlying technical mechanism(s) or protocol(s). We also identified the core purpose of each app and grouped apps with similar functionalities to categorize the landscape of government-backed corona apps.
}

\changed{
To reduce complexity for our study, we streamlined the CI-based aspects as follows: We did not differentiate between primary and secondary data receivers (respectively, the organization responsible for the data processing and third parties the data could possibly be shared with). Similarly, we chose not to consider multi-stage data transmissions (\eg, encounter information initially being sent to no one and only stored on the phone but transmitted to health authorities' servers after a positive test result had been registered).
To reduce cognitive load for our participants, we differentiated between payload data and to what degree it allowed for identification of the app user.
This left us with five CI-based data collection \emph{factors} (2.--6. in the list below), to which we added the app's purpose and the aforementioned societal and technical implications.

For each factor, we determined a set of concrete \emph{factor levels} which we would later combine into scenarios of hypothetical yet realistic corona apps for our survey.
The factor levels were derived from the respective information collected from real-world corona apps, with the goal of reflecting the range of real-world practices while keeping the number of factor levels as small as possible.
For example, the observed data retention times could be classified as fixed short-term intervals (between 14 days, \ie, the typical duration of quarantine for COVID-19 patients, and several weeks), unspecified but identifiable points in time (``when the need for contact tracing is over''), and unspecified / indefinite. This spectrum is reflected in the final set of factor levels: one month (fixed short-term interval), until the end of the current coronavirus regulations (unspecified but identifiable), unspecified.
}

The final sets of factors and factor levels are as follows:

\begin{enumerate}
    \setlength{\itemsep}{0pt}
    \setlength{\parskip}{0pt}
    \setlength{\parsep}{0pt}
    \item \textbf{Purpose (5 levels):} contact tracing, symptom check, quarantine enforcement, information, health certificate.
    \item \textbf{Data Collected (16 levels):} encounter data, location data, health or activity data (excluding {COVID-19} infection status), {COVID-19} infection status, all combinations of two or three of the aforementioned items\footnote{We omitted the combination of all four items because this would have made the scenario descriptions too long.}, unspecified data (``data from your smartphone''), no data. \changed{For this factor we allowed combinations of factor levels to occur because otherwise certain app functionalities would not have made sense (\eg, digital contact tracing can only work if a person's encounter data and infection status are provided).}
    \item \textbf{User Anonymity (3 levels):} whether the app collects personal data that allows for unique identification of the individual, collects only demographic data, or collects only data that cannot be used to uniquely identify the user.
    \item \textbf{Data Receiver (6 levels):} health authorities, law enforcement, research institutions, private companies, the public, none.
    \item \textbf{Data Transmission (3 levels):} automatically, manually (app-type-specific wording, \eg, for symptom check: ``when a symptom check is requested''), none (in case no data is collected).
    \item \textbf{Retention (3 levels):} one month, until end of current coronavirus regulations, unspecified.
    \item \textbf{Technical Implications (3 levels):} impact on battery life, app malfunctioning (app-type-specific wording, \eg, false positive for breaking quarantine), none.
    \item \textbf{Soci(et)al Implications (4 levels):} possible additional benefits in the future, more timely adjustment of local coronavirus regulations, extended personal freedom of movement or travel, none.
\end{enumerate}

\begin{figure*}[htbp]
\Description[Example vignette with highlighted text passages]{
The example vignette figure shows the text of a specific scenario. The figure highlights which text passages correspond to which factor. The text is as follow (with highlight tags):
Imagine an app [Purpose]that provides information about your health and needs to be shown if you want to visit a certain place.[/Purpose]
The app uses [Data]health or activity data, your COVID-19 infection status, and your current or past location(s).[/Data]
[User Anonymity]In addition, the app collects data that could be used to uniquely identify you.[/User Anonymity]
This data [Receiver]is sent to research institutions[/Receiver] [Transmission]when you request your health report and[/Transmission] [Retention]it will be stored until the current coronavirus regulations end[/Retention]
[Technical Implication]The app decreases your phone’s battery life.[/Technical Implication]
[Societal Implication]Using this app may increase your personal freedom of movement or travel.[/Societal Implication]
}
\centering
\includegraphics[width=\linewidth]{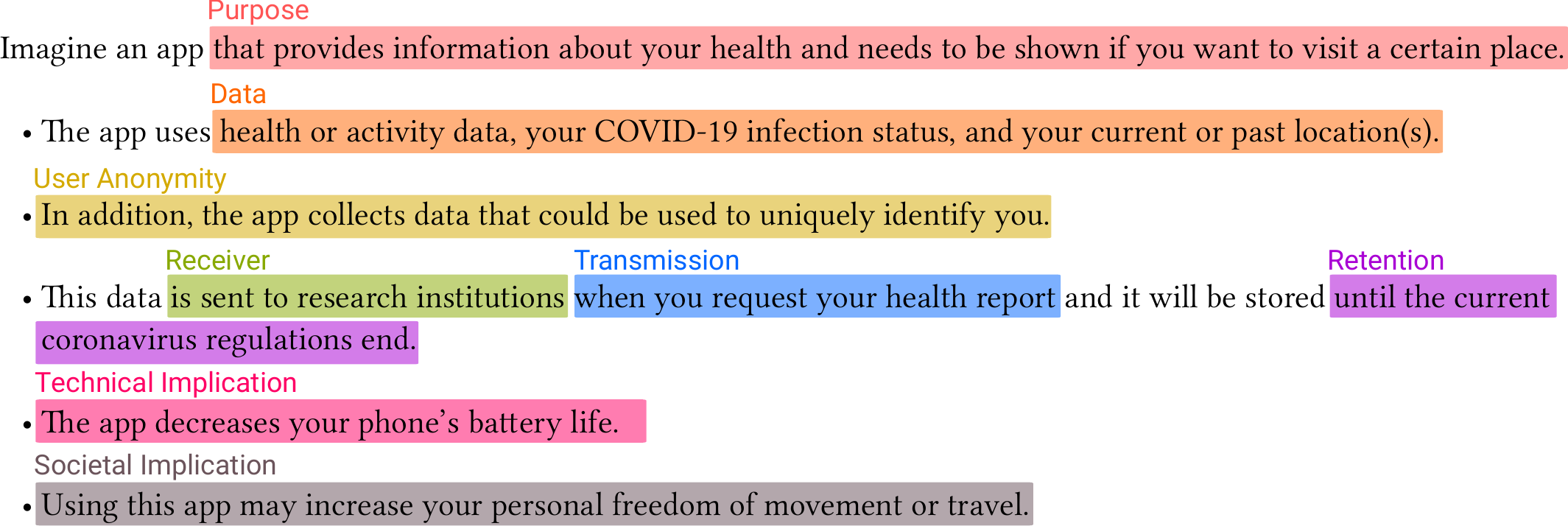}
\caption{\label{fig:scenario_example}Example of a vignette that combines different \emph{factor levels} into a specific \emph{scenario}.}
\end{figure*}

\subsubsection{Vignette Composition}
\label{sec:vignette_design}
In our survey a \emph{vignette}, as shown in Figure~\ref{fig:scenario_example}, is a short description of a hypothetical smartphone app, consisting of an immutable text template (the non-highlighted black text) with placeholders for the eight \emph{factors} (colored boxes), each of which was assigned one out of multiple different \emph{factor levels} (the text in the colored boxes) to create a specific \emph{scenario}.
Across different scenarios and participants, we systematically altered the factor levels in order to measure how the individual factors and factor levels affected our participants' assessment of the hypothetical apps.

Starting from the set of all \num{15\,5520} possible combinations of factor levels that could theoretically compose a scenario, we carefully defined \emph{dependencies} between specific factor levels in order to exclude scenarios whose combinations of factor levels would not make sense.
The specification of dependencies was the outcome of thorough discussions between three researchers and each feature dependency in the set required joint agreement.

\changed{
The following list describes the dependencies between different factor levels for the composition of vignettes.
\begin{itemize}
    \setlength{\itemsep}{0pt}
    \setlength{\parskip}{0pt}
    \setlength{\parsep}{0pt}
    \item Symptom check apps always required health or activity data.
    \item Contact tracing apps always required the infection status and either encounter data, location data, or both.
    \item Quarantine enforcement apps always required location data and unique identification of the individual user.
    \item Health certificate apps always required the COVID-19 infection status and the societal implication of extended personal freedom of movement or travel.
    \item Whenever the use of an app implied as societal implication extended personal freedom of movement or travel, it was required that the data collected by the app allowed for unique identification of the individual.
    \item For data made available to the public, data retention was set to ``unspecified''.
\end{itemize}

Applying these dependencies to our initial set of possible scenarios resulted in a final set of \num{50\,625} unique vignettes, which were distributed across the \num{5} app purposes as follows: \num{14\,850} for contact tracing, \num{9450} for symptom check, \num{3780} for quarantine enforcement, \num{21\,600} for information, and \num{945} for health certificate.
}

\changed{
From this main set of unique vignettes, we composed individual sets of \num{10} vignettes for each questionnaire.
For each participant, we (i) randomly selected two vignettes per app purpose under the condition that (ii) these two vignettes were different in every factor level (as long as this condition could be fulfilled).
Within a participant's set, vignettes were always shown in random order.
}

\subsection{Questionnaire}

The vignettes constituted the core of our questionnaire, which we outline in the following. The full version can be found in Appendix~\ref{appendix:questionnaire}.

\subsubsection{Introduction}
At the beginning of the questionnaire, we introduced the purpose of the study, provided information about data collection and processing, and asked for the participants' consent to proceed. \changed{Section~\ref{sec:ethics} provides further information how we protected participants' privacy.}

\subsubsection{Smartphone Use}
Possessing a smartphone is crucial to determine how the use of mobile apps is perceived by participants.
Therefore, we asked if participants owned a smartphone~(Q1) and, if yes, what operating system they used~(Q2) and how satisfied they were with certain aspects (\eg, battery life) of their smartphone~(Q3).

\subsubsection{Context: Coronavirus}
Personal experience with the coronavirus was found to be a significant factor in the adoption of (hypothetical) contact tracing apps~\cite{zhang_covid_privacy_2020}. Hence, after defining key terms, we asked whether participants~(Q4) or people in their social circles~(Q5) had tested positive for coronavirus and if participants had been quarantined or quarantined themselves because of the virus~(Q6).
We further asked if they lived with a person at higher risk~(Q7) and how concerned they were about someone close to them becoming infected with the coronavirus~(Q8).

\subsubsection{App Scenarios}
As described in Section~\ref{sec:vignette_design}, we presented each participant with a unique set of ten (hypothetical) apps. For each app, we asked participants how likely they were to use the described app~(Q9) and to estimate the expected amount of app users in their country~(Q10).
We further asked them about the perceived normative pressure Ajzen \cite{ajzen_tbpquestionnaire_2019} in their social circles to use the app~(Q11) and to assess the usefulness of the app to fight the global pandemic~(Q12).

\subsubsection{Experience with Corona Apps}
Next, we asked participants if they were aware of existing corona apps in their country for any of the five purposes covered in this study~(Q13).
We further asked them to indicate if they were actively using such an app~(Q14) and, if yes, which one~(Q14a), or why they did not do so in the opposite case~(Q14b).
At the end of this section, participants were asked the open-ended questions of potential positive~(Q15) and negative~(Q16) aspects of corona apps.

\subsubsection{Attitudes Towards Governmental Actions}
To assess participants' general attitude towards the institutions tasked with fighting the pandemic, we asked three questions:
First, we asked participants to rate the measures applied in their region to counter the pandemic~(Q17). 
We further asked them about their opinion of six public institutions: health authorities, law enforcement, research institutions, private companies, and federal and regional governments~(Q18).
This part concluded with a question about the acceptance of private companies' practice to share anonymized data, such as aggregated phone location data, with public authorities to help limit the spread of the pandemic~(Q19), which is a practice used in several countries such as the US and Germany~\cite{telekom_prediction_2020}.

\subsubsection{Individual Privacy Concerns}
Finally, we measured participants' individual privacy concerns using the Internet Users' Information Privacy Concerns (IUIPC) constructs for Control, Awareness (of privacy practices), and Collection~\cite{malhotra_iuipc_2004}.

\subsection{Data Collection}
\label{sec:data-collection}

We implemented an English-language preliminary version of our survey to ensure comprehensibility and estimate timings. The pre-study survey was
distributed to experienced researchers and within our own social circles via snowball sampling.
33 participants completed the survey between June~1 and June~5, 2020. Based on participants' feedback, we changed the presentation of the scenarios to improve readability and highlight differences between the different apps. 

In the period from June to August 2020, the online survey was distributed in three different countries: Germany, the United States, and China.
These three countries were selected for their different approaches to corona apps and for their important roles in the global pandemic: China as the country which was struck first, the USA, which quickly became the most affected country in terms of confirmed cases, and Germany as a representative for a European country that handled the first wave of the pandemic relatively well.
At the time of the survey all countries were in different states of the pandemic, which we outline in the following subsections to provide context for our results.
\changed{
We purchased the online panels as a full service from Lightspeed Research (Kantar), including survey implementation and translation, participant recruitment, and data quality assurance (cost was EUR~$2500$ per online panel plus EUR~$4000$ for implementation and translation).
Representativity quotas were matched with an average discrepancy of \num{2.5}\,\% (US), \num{2}\,\% (Germany), and \num{2.9}\,\% (China).
Respondents faster than 40\% of the median response time were directly discarded by Kantar.

We created the German and English versions of the questionnaire ourselves.
The translation of the English questionnaire into Mandarin and the translation of the Chinese open-ended answers into English were commissioned to a translation agency. 
With the help of a native speaker from our social circles, we followed the quality control process of back translation for the questionnaire and corrected a few minor errors.
Back translation of the open-ended questions (Q14a, Q14b, Q15, Q16) was carried out on several random samples as the cost of a complete back translation did not seem to justify the expected benefit.
}

\subsubsection{Germany}
\label{subsubsec:germany}
The study was conducted surveying a sample of \num{1003} participants representative for the German adult population by gender, age, region, and education from June~9 to June~11, shortly before the launch of the national contact tracing app, ``Corona-Warn-App'' \cite{sap_coronawarnapp_2020}, on June~16.
This app uses the Exposure Notification API~\cite{applegoogle_ppct_2020}, which is influenced by the Decentralized Privacy-Preserving Proximity Tracing (DP-3T)~\cite{troncoso_dp3t_2020} and the Temporary Contact Number (TCN)~\cite{tcncoalition_tcn_2020} protocols.
At the time of the study, two other corona apps supported by the federal government were already available in Germany:
``NINA'' is the government's general emergency information app currently used to provide information about regional coronavirus regulations and about the state of and recommended behavior during the pandemic (available since June~1, 2015; coronavirus reports since March 2020).
The ``Corona-Datenspende'' \cite{rki_coronadatenspende_2020} app is a research project by the Robert Koch Institute (RKI, German federal government agency responsible for disease control and prevention), which collects and analyzes vital data recorded by fitness trackers to help assess the number of people with COVID-19 (available since April~7, 2020).

Germany reported its first SARS-CoV-2 cases in February, with the highest number of new cases in late March and early April. The spread of the virus was primarily fought through strict contact restrictions, which were in effect from March~22 to early May. Public debate about apps for digital contact tracing started in April and received media attention over the following weeks including discussions about data protection and the underlying protocols.
As of June~9, 2020, when the survey was distributed, Germany had \num{18\,4543} confirmed cases of infected people, thereof \num{17\,0200} estimated recoveries, \num{5632} estimated active cases, and \num{8711} deaths~\cite{rki_covid19dashboard_2020}.
The national 7-day rolling average of daily new cases per \num{10\,0000} residents was very low at \num{3.0}, with thresholds of \num{35} and \num{50} on county level for tightening the measures to fight the pandemic.

\subsubsection{United States}
\label{subsubsec:us}
In the US, we surveyed a sample of \num{1003} participants representative for the US adult population by gender, age, region, and education from July~6 to July~14.
At the time of the study, one corona app supported by the federal government was already available:
The ``COVID-19 Screening Tool''~\cite{apple_covidscreening_2020} is an app and web application made available on March~27, which recommends actions based on user-provided information such as symptoms, contacts, and travel. It was developed by Apple and the Centers for Disease Control and Prevention~(CDC).
While several public and private initiatives had been developing digital contact tracing apps for the US, no nation-wide rollout was expected for the near future. At the time of writing, six US states and Guam had released apps based on Apple's and Google's Exposure Notification API~\cite{sholtz_tracingapp_2020}.

The United States were hit by the pandemic in mid-March and soon became the country with the highest number of officially confirmed cases worldwide.
Stay-at-home orders came into effect in 44 states between March~20 and April~7, 2020 and were lifted between April 26 and June~11, 2020.
However, the number of cases increased again during July and August.
As of July~6, 2020, when the survey was distributed, the US had \num{2893083} officially confirmed cases of infections, thereof \num{13\,24947} estimated recoveries, \num{14\,48438} estimated active cases, and \num{11\,9698} deaths~\cite{cdc_coviddatatracker_2020}.
The national 7-day rolling average of daily new cases per \num{10\,0000} residents was at \num{103.5}.

\subsubsection{China}
\label{subsubsec:china}
In China, we surveyed a sample of \num{1019} participants representative for the Chinese adult \emph{online} population\footnote{Due to the nature of this online study certain predominantly rural population groups could not be surveyed.} by gender and age from July~27 to August~6.
In February, China was the first country to introduce various mandatory health code systems at the local and regional level.
These systems use QR codes that need to be scanned with mobile applications to track and ultimately prevent the movement of potentially infected individuals based on basic health information and travel history~\cite{mozur_china_2020}.
The existing health code systems serve multiple purposes, such as providing general information about the coronavirus, calculating the risk of infection, contact tracing, and quarantine enforcement.
In contrast to Germany and the US, most Chinese ``corona apps'' are integrated into widely used platform economies, \eg, WeChat or Alipay, which fostered their quick adoption.

China's Hubei province was the epicenter of the worldwide SARS-CoV-2 outbreak in January and February 2020.
The central government reacted with several strict regional lockdowns, effectively quarantining more than 60~million people (\eg, Wuhan, Hubei, January 23 to April 8).
The exponential growth of officially confirmed cases in China ended in early March.
As of July~27, 2020, China had \num{83\,891} officially confirmed cases of infected people, thereof \num{78\,918} estimated recoveries, \num{339} estimated active cases, and \num{4634} deaths~ \cite{nhc_dailybriefing_2020}.
The national 7-day rolling average of daily new cases per \num{10\,0000} residents was at \num{0.02} with only \num{209} new cases all over China between July~21 and July~27.

\changed{
\subsection{Research Ethics}        
\label{sec:ethics}
Our department does not have an institutional review board. Instead, our study followed best practices of human subject research and data protection guidelines, including the rules of the European GDPR.
All data protection measures were reviewed and approved by our institution's data protection office.
Kantar, our panel provider, has commited itself to follow the ICC/ESOMAR code of conduct~\cite{ICCESOMAR_Code_2021}.
}

\subsection{Data Analysis}
\subsubsection{Statistical Analysis}
\label{subsubsec:method:statistic}

To understand which factors influence participants' decisions to use an app (Q9 = outcome variable), we performed a regression analysis for the data utilizing the cumulative link models module of the \texttt{ordinal} R package~\cite{christensen_cumulative_2018}.
In line with best practice~\cite{zuur_mixed_2009}, we first calculated a model containing all factors available in our data set and successively removed non-significant factors based on the Akaike information criterion (AIC) of the resulting model to determine the model that best fit the data. We conducted this analysis for each country individually as the circumstances and timing of the data collection in each country varied (see Section~\ref{sec:data-collection}).

For each factor we chose as a baseline the level that we expected to have the least effect (\eg, the ``information'' purpose or the middle choice for Likert-scale questions). For some of the Likert-scale questions we grouped the answers to reduce the number of factor levels in the model, \eg, the assessments of the government response~(Q17) ``too lenient'' and ``way too lenient'' were combined into ''too lenient''. We also excluded all responses that chose ``prefer not to answer''.

\subsubsection{Qualitative Analysis}

To analyze the open-ended answers to the questions why participants did not use a corona app~(Q14a) and what they perceived to be positive~(Q15) and negative~(Q16) aspects of corona apps, we used an iterative open coding procedure. \changed{We used a mixed-methods approach with qualitative and quantitative elements, as suggested by Mayring~\cite{mayring_qualitative_2014}. We first assigned codes to the text (inductive categorization) and analyzed the frequency of the codes afterwards.} Coding was performed by two researchers with interdisciplinary backgrounds in psychology and information security. In a first step, the two coders independently coded the first 300 open responses from each country to derive common themes and create an initial coding frame for each question. This process consisted of closely examining participants' answers to identify and conceptualize categories. Each item could be assigned one or multiple codes.

The coders then discussed their codes and agreed on a final coding frame for each question. The frame was validated by both coders coding answers 301--500 for each question (20\,\% of the data, which is in the typical 10--25\,\% range to determine coder agreement~\cite{oconnor_icr_2020}). As a measure for inter-coder reliability, ReCal2~\cite{freelon_recal_2010} was used to compute Krippendorff's alpha for each code, of which we report a weighted mean that takes into account the frequency of each code. Finally, the remaining sets of 500 answers for each question and country were coded by a single coder.

\section{Results}

In this section, we first describe our samples based on demographic data and participants' survey responses.
We then compare the responses to the four questions asked for each scenario and analyze the effect sizes of several factors that impact users' willingness to use corona apps across the three countries. Finally, we look in detail at participants' individual perceptions of corona apps.

\begin{table*}[htbp]
\Description[Participants' demographics table]{
The demographics of the participants table consists of eight columns: a column for the row group label, the row label, and six data columns, two for each country. The table header labels the data columns with Germany, United States, and China. It also contains the number of participants for each country. The table rows are grouped in the seven categories gender, age, education, smartphone use, virus experience, IUIPC, and opinion of, where each section spans multiple rows. Each category has a sub-header for the data columns which is either number of participants and percentage per country or mean and standard derivation per country.
}

\centering
\caption[Participant Demographics]{\label{tab:demographics}
Participant Demographics. Data for gender, age, and education as delivered by our panel provider. Information about participants' smartphone use, previous coronavirus experience, and general privacy attitudes was collected in the questionnaire.
For satisfaction with certain aspects of their smartphones and privacy attitudes we provide average response values (mean and sd) [smartphone satisfaction measured with response scales from \num{1} (\emph{very satisfied}) to \num{5}; privacy attitudes measured with Likert agreement scales from \num{1} (\emph{strongly disagree}) to \num{7}, trust in institutions measured with Likert scales from \num{1} (\emph{very comfortable}) to \num{5}].
}
\begin{minipage}{\textwidth}
\renewcommand{\thefootnote}{\thempfootnote}
\renewcommand{\arraystretch}{0.75}
\begin{tabular*}{\textwidth}{@{}>{\bfseries}ll@{\extracolsep{\fill}}*{6}{r}}
\toprule
&                                    & \multicolumn{2}{c}{\textbf{Germany}}        & \multicolumn{2}{c}{\textbf{United States}}  & \multicolumn{2}{c}{\textbf{China}}          \\
&                                    & \multicolumn{2}{c}{\emph{(n = \num{1003})}} & \multicolumn{2}{c}{\emph{(n = \num{1003})}} & \multicolumn{2}{c}{\emph{(n = \num{1019})}} \\

\midrule\multirow{3}{*}{\rotatebox[origin=c]{90}{Gen.}}
& & \multicolumn{1}{c}{\textbf{n}} & \multicolumn{1}{c}{\textbf{\%}} & \multicolumn{1}{c}{\textbf{n}} & \multicolumn{1}{c}{\textbf{\%}} & \multicolumn{1}{c}{\textbf{n}} & \multicolumn{1}{c}{\textbf{\%}} \\
& Female                            &  508    & 50.7  &  532    &  53.0  &   495    &  48.6  \\
& Male                              &  495    & 49.4  &  471    &  47.0  &   524    &  51.4  \\
\midrule\multirow{5}{*}{\rotatebox[origin=c]{90}{Age}}
& & \multicolumn{1}{c}{\textbf{n}} & \multicolumn{1}{c}{\textbf{\%}} & \multicolumn{1}{c}{\textbf{n}} & \multicolumn{1}{c}{\textbf{\%}} & \multicolumn{1}{c}{\textbf{n}} & \multicolumn{1}{c}{\textbf{\%}} \\
& 18--35                            &  262    & 26.1  &  199    &  19.8  &   506    &  49.7  \\
& 36--50                            &  246    & 24.5  &  297    &  29.6  &   312    &  30.6  \\
& 51--65                            &  374    & 37.3  &  302    &  30.1  &   165    &  16.2  \\
& 66--80                            &  121    & 12.0  &  205    &  20.4  &    36    &   3.5  \\
\midrule\multirow{6}{*}{\rotatebox[origin=c]{90}{Education}}
& & \multicolumn{1}{c}{\textbf{n}} & \multicolumn{1}{c}{\textbf{\%}} & \multicolumn{1}{c}{\textbf{n}} & \multicolumn{1}{c}{\textbf{\%}} & \multicolumn{1}{c}{\textbf{n}} & \multicolumn{1}{c}{\textbf{\%}} \\
& Less than high school              &  162    & 16.2  &   74  &   7.4  &    16  &   1.6  \\
& High school or associate degree    &  602    & 60.0  &  407  &  40.6  &    99  &   9.7  \\
& Undergraduate degree\footnote{No distinction between undergraduate and (post)graduate degrees in Germany as this was only introduced in the 2000s.} & {\multirow{2}{*}{239}} & \multirow{2}{*}{23.8} & 425 & 42.4  &   470 & 46.1  \\
& Postgraduate degree\footnotemark[\value{mpfootnote}] &         &       &   97  &   9.7  &   103  &  10.1  \\
& No data                            &    0    &  0.0  &    0  &   0.0  &   331  &  32.5  \\
\midrule\multirow{9}{*}{\rotatebox[origin=c]{90}{\parbox{60pt}{\centering Smartphone Use}}}
& & \multicolumn{1}{c}{\textbf{n}} & \multicolumn{1}{c}{\textbf{\%}} & \multicolumn{1}{c}{\textbf{n}} & \multicolumn{1}{c}{\textbf{\%}} & \multicolumn{1}{c}{\textbf{n}} & \multicolumn{1}{c}{\textbf{\%}} \\
& Phone owners                       &  934    & 93.1  &  873  &  87.0  &  1009  &  99.0  \\
& Corona app users                   &   42    &  4.2  &   66  &   6.6  &   614  &  60.3  \\
\addlinespace
& & \multicolumn{1}{c}{\textbf{mean}} & \multicolumn{1}{c}{\textbf{sd}} & \multicolumn{1}{c}{\textbf{mean}} & \multicolumn{1}{c}{\textbf{sd}} & \multicolumn{1}{c}{\textbf{mean}} & \multicolumn{1}{c}{\textbf{sd}} \\
& Battery satisfaction               &  2.12   &  0.92 &  2.11 &   1.00 &   2.10 &   0.98 \\
& Location satisfaction              &  1.91   &  0.72 &  1.84 &   0.76 &   1.82 &   0.74 \\
& Camera satisfaction                &  1.89   &  0.85 &  1.88 &   0.87 &   1.85 &   0.77 \\
& Speed satisfaction                 &  1.92   &  0.77 &  1.93 &   0.82 &   1.97 &   0.86 \\
\midrule\multirow{6}{*}{\rotatebox[origin=c]{90}{\parbox{50pt}{\centering Virus Experience}}}
& & \multicolumn{1}{c}{\textbf{n}} & \multicolumn{1}{c}{\textbf{\%}} & \multicolumn{1}{c}{\textbf{n}} & \multicolumn{1}{c}{\textbf{\%}} & \multicolumn{1}{c}{\textbf{n}} & \multicolumn{1}{c}{\textbf{\%}} \\
& Tested positively                  &   13    &  1.3  &   22  &   2.2  &    18  &   1.8  \\
& Infection in social circles        &  101    & 10.1  &  135  &  13.5  &   100  &   9.8  \\
& Quarantine experience              &  147    & 14.7  &  367  &  36.6  &   293  &  28.8  \\
& Risk person in household           &  478    & 47.7  &  450  &  44.9  &   375  &  36.8  \\
& Infection concerns                 &  112    & 11.2  &  124  &  12.4  &    91  &   8.9  \\
\midrule\multirow{7}{*}{\rotatebox[origin=c]{90}{Opinion of}}
& & \multicolumn{1}{c}{\textbf{mean}} & \multicolumn{1}{c}{\textbf{sd}} & \multicolumn{1}{c}{\textbf{mean}} & \multicolumn{1}{c}{\textbf{sd}} & \multicolumn{1}{c}{\textbf{mean}} & \multicolumn{1}{c}{\textbf{sd}} \\
& Health authorities                 &  2.45   &  0.89 &  2.40 &   1.01 &   1.75 &   0.76 \\
& Law enforcement                    &  2.57   &  0.92 &  2.62 &   1.05 &   1.83 &   0.91 \\
& Research institutions              &  2.31   &  0.95 &  2.38 &   0.98 &   1.86 &   0.90 \\
& Private companies                  &  2.89   &  0.80 &  2.82 &   0.92 &   2.27 &   0.88 \\
& Federal government                 &  2.56   &  1.14 &  3.28 &   1.27 &   1.52 &   0.80 \\
& State government                   &  2.61   &  1.11 &  2.77 &   1.19 &   1.73 &   0.80 \\
\midrule\multirow{6}{*}{\rotatebox[origin=c]{90}{\parbox{50pt}{\centering Known Apps}}}
& & \multicolumn{1}{c}{\textbf{n}} & \multicolumn{1}{c}{\textbf{\%}} & \multicolumn{1}{c}{\textbf{n}} & \multicolumn{1}{c}{\textbf{\%}} & \multicolumn{1}{c}{\textbf{n}} & \multicolumn{1}{c}{\textbf{\%}} \\
& Info                               &  199    & 19.8  &  154  &  15.4  &   723  &  71.0  \\
& Symptom check                      &  158    & 15.8  &  150  &  15.0  &   448  &  44.0  \\
& Quarantine enforcement             &  122    & 12.2  &   98  &   9.8  &   443  &  43.5  \\
& Contact tracing                    &  208    & 20.7  &  127  &  12.7  &   527  &  51.7  \\
& Health certificate                 &  124    & 12.4  &  122  &  12.2  &   783  &  76.8  \\
\midrule\multirow{4}{*}{\rotatebox[origin=c]{90}{IUIPC}}
& & \multicolumn{1}{c}{\textbf{mean}} & \multicolumn{1}{c}{\textbf{sd}} & \multicolumn{1}{c}{\textbf{mean}} & \multicolumn{1}{c}{\textbf{sd}} & \multicolumn{1}{c}{\textbf{mean}} & \multicolumn{1}{c}{\textbf{sd}} \\
& Control                            &  6.16   &  1.20 &  5.51 &   1.24 &   5.46 &   0.93 \\
& Awareness                          &  5.43   &  1.32 &  6.04 &   1.18 &   5.69 &   0.97 \\
& Collection                         &  5.68   &  1.17 &  5.64 &   1.29 &   5.32 &   1.05 \\
\bottomrule
\end{tabular*}
\end{minipage}
\end{table*}

\subsection{Sample Description}

In Table~\ref{tab:demographics}, we provide demographic information (gender, age, and education) about our participants as delivered by our panel provider and additional information collected in our survey.

For all three countries, we observe higher smartphone prevalence than reported in recent data~\cite{statista_smartphoneownership_2019}, but our numbers are comparable to those of smartphones and feature phones combined.
The high prevalence (\SI{99}{\percent}) of smartphone use in China reflects that our sample is representative for the more technophile online population (see Section~\ref{subsubsec:china}).

Although participants seemed to be rather satisfied with their current phones, they were, across all countries, significantly less satisfied with their phones' battery life than with other properties.

Between \SI{1.3}{\percent} and \SI{2.2}{\percent} of participants reported that they had tested positive for coronavirus. 
While these numbers are higher than those of officially confirmed cases for each country (see Section~\ref{sec:data-collection}), they fall within the margin of error of \SI{3}{\percent}.
In all countries, about \SI{10}{\percent} of participants reported that a person in their social circles had been infected with the virus.
Between countries, we observe differences in the number of participants who had been under quarantine since the beginning of the pandemic, with the US reaching the highest rate (\SI{37}{\percent}) and Germany the lowest (\SI{15}{\percent}).
Despite different states of the pandemic in the surveyed countries, similar numbers of participants (about \SI{10}{\percent}) were concerned about becoming infected with coronavirus.

Table~\ref{tab:demographics} also provides information about participants' opinions of institutions tasked with measures to counter the pandemic, what types of corona apps participants already knew to exist, and their attitude towards online privacy using three IUIPC dimensions.

\subsection{Cross-Country Assessment of App Scenarios}

Next, we compare the responses to the four questions asked for each hypothetical app scenario (Q9~to~Q12) across the three countries.
Table~\ref{tab:scenario-descriptives} shows mean response values per country along with standard deviations across all app scenarios.
Questions were answered on numeric 7-point scales with a higher number representing a more positive response, \ie, in the case of Q9, higher willingness to use the presented app. 
Participants in China were generally more willing to use the apps presented in the scenarios, with Germany ranked in the middle and US participants expressing the lowest average willingness to use an app.
With a one-way analysis of variance (ANOVA) we found statistically significant differences ($p < .05$) for Q9 between all three countries.
Post-hoc $t$-tests of independence also showed that individual differences between all pairs of countries were significant ($p < .05$ equivalent, values Bonferroni-corrected for multiple testing).
Expected adoption~(Q10), perceived pressure~(Q11), and perceived utility~(Q12) of corona apps were significantly higher in China than in Germany and the US.
We found no significant differences for Q10, Q11, and Q12 between Germany and the US.

\begin{table*}[htbp]
\Description[Descriptive statistic table of the responses to questions Q9 to Q12]{
The descriptive statistic table of the responses to questions Q9 to Q12 consists of eight columns: a column containing the question code, the question name, and six data columns, two for each country. The table header labels the data columns with Germany, United States, and China. Each row contains the mean and standard derivation per country for the questions Q9, Q10, Q11, and Q12.
}
\centering
\caption[Average response values (mean and sd) for the four scenario questions (Q9~to~Q12)]{
Average response values (mean and sd) for the four scenario questions (Q9~to~Q12), 7-point scales with higher numbers indicating a more positive response.
\label{tab:scenario-descriptives}
}
\begin{tabular*}{\textwidth}{@{}>{\bfseries}ll@{\extracolsep{\fill}}*{6}{c}}
\toprule
&  & \multicolumn{2}{c}{\textbf{Germany}} & \multicolumn{2}{c}{\textbf{United States}} & \multicolumn{2}{c}{\textbf{China}} \\
\midrule
& & \textbf{mean} & \textbf{sd} & \textbf{mean} & \textbf{sd} & \textbf{mean} & \textbf{sd} \\
Q9  & Willingness to use & 3.25 & 2.04 & 3.12 & 2.10 & 5.27 & 1.55 \\
Q10 & Expected adoption  & 3.43 & 1.52 & 3.43 & 1.72 & 5.06 & 1.38 \\
Q11 & Perceived pressure & 3.38 & 1.86 & 3.32 & 1.96 & 5.19 & 1.50 \\
Q12 & Perceived utility  & 3.69 & 1.89 & 3.62 & 1.98 & 5.37 & 1.44 \\
\bottomrule
\end{tabular*}
\end{table*}

Table~\ref{tab:q9_matrix} takes a closer look at participants' willingness to use corona apps (Q9) by country and app purpose. The figures indicate how many participants tended towards a positive answer (above the medium response value, \ie,~\num{4}) regarding their willingness to use the presented app. Within countries, we observe only small differences in the numbers for different app purposes, while a cross-country comparison yields the general willingness to use corona apps for any purpose to be much higher in China compared to Germany and the US.
Across all countries, we observe the highest willingness to use for contact tracing apps and the lowest for quarantine enforcement.
However, these values can only provide a first impression since they represent averages per app type and country and contain interfering effects between all factors that constitute a scenario.
Section~\ref{sec:clmm} studies the individual effects in more detail.

\begin{table*}[htbp]
\Description[Descriptive statistic table of participants to use an app]{
The descriptive statistic table of participants to use an app consists of 6 columns: a row label column and five app purpose columns. Table header labels are the five app purposes, e.g., symptom check or health certificate. There are three rows, one for each country. Each row shows one percentage per purpose.
}
\centering
\caption[Percentages of participants that are (very) likely to use an app for the given purpose.]{\label{tab:q9_matrix}
Percentages of participants that are (very) likely to use an app for the given purpose, regardless of the specifics.
}
\begin{tabular*}{\textwidth}{@{}>{\bfseries}r@{\extracolsep{\fill}}*{5}{c}}
\toprule
& \textbf{Symptom Check} & \textbf{Contact Tracing} & \textbf{Quarantine Enf.} & \textbf{Information} & \textbf{Health Certificate} \\
\midrule
Germany       & 32\,\% & 37\,\% & 27\,\% & 30\,\% & 26\,\% \\ 
United States & 29\,\% & 32\,\% & 25\,\% & 26\,\% & 27\,\% \\
China         & 74\,\% & 80\,\% & 72\,\% & 77\,\% & 76\,\% \\
\bottomrule
\end{tabular*}
\end{table*}

\subsection{Factors Impacting the Willingness to Use a Corona App}
\label{sec:clmm}
Table~\ref{tab:clmm_results_vignettes} lists the data collection factors and factor levels in our app scenarios that influence participants’ reported willingness to use a corona app, along with their respective effect sizes and $p$-values.
Responses collected in the non-scenario parts of our survey that we found to impact participants' willingness to use a corona app are listed in Appendix~\ref{appendix:clmm_country_overview}, Tables~\ref{tab:clmm_results} and~\ref{tab:clmm_results2}.
Statistically significant factor levels and answers ($p < .05$) are printed in bold in all tables.
Positive estimate values indicate a positive influence on the willingness to use an app, negative estimates indicate tendencies to not use an app. During the model optimization process several factors were excluded in some countries while remaining in the model for the other countries (see Section~\ref{subsubsec:method:statistic} and ~\ref{appendix:clmm_country_overview}). For example, the societal benefit on the regional level (see Table~\ref{tab:clmm_results_vignettes}) is a significant predicting factor in the US and Germany but did not make it into the final model for China.
In the following, we report for each country which factor levels and non-scenario answers positively or negatively influenced participants' decision to use an app and touch upon the differences between the three countries. 

\begin{table*}[htbp]
\Description[Scenario-based CLMM regression table of participants' willingness to use corona apps]{
The scenario-based CLMM regression table consists of eight columns: a column to label the factor and the factor level in each row, a column depicting the effect size as bar chart (one bar for each country), and six data columns---two for each country. The data columns form three pairs of an estimate and a p-value column. The table header labels reads as ``Factor'', ``+/-'' (increase or decrease), and the country names where each country has a sub header ``Estimate'' and ``Pr(>|z|)''. The rows are grouped by eight factors and the first row in each of these groups contains only the factor name and the baseline factor level. The other rows contain the factor levels and the corresponding data.
}
\centering
\caption[CLMM regression of participants' willingness to use corona apps. (Summary)]{\label{tab:clmm_results_vignettes}
Cumulative link mixed model for participants' willingness to use corona apps. A positive estimate (effect size) indicates participants' willingness to use the app being higher compared to the factor's baseline variable. All factor levels in this table were incorporated into the app scenarios (vignettes). Only factors present in the final model are presented here. The colored bars represent the effect sizes per country (yellow: Germany, blue: United States, red: China).
}
\begin{tabular*}{\textwidth}{@{}ll@{\extracolsep{\fill}}*{6}{r}@{}}
\toprule
Factors and Levels & \multicolumn{1}{c}{+/-} & \multicolumn{2}{c}{\textbf{Germany}} & \multicolumn{2}{c}{\textbf{United States}} & \multicolumn{2}{c}{\textbf{China}} \\
&& Estimate & Pr(>|z|) & Estimate & Pr(>|z|) & Estimate & Pr(>|z|) \\

\midrule\multicolumn{8}{l}{\emph{Purpose (baseline: information)}}\\
 Symptom check & \includegraphics[page=1]{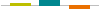} & 0.06 & 0.21 & \textbf{0.19} & \textbf{< 0.01} & \textbf{-0.12} & \textbf{0.03}\\
 Contact tracing & \includegraphics[page=2]{figures/tab_clmm_results_vignettes.pdf} & \textbf{0.38} & \textbf{< 0.01} & \textbf{0.42} & \textbf{< 0.01} & 0.09 & 0.08\\
 Quarantine enforcement & \includegraphics[page=3]{figures/tab_clmm_results_vignettes.pdf} & -0.08 & 0.19 & -0.03 & 0.58 & \textbf{-0.28} & \textbf{< 0.01}\\
 Health certificate & \includegraphics[page=4]{figures/tab_clmm_results_vignettes.pdf} & -0.11 & 0.09 & \textbf{0.21} & \textbf{< 0.01} & -0.07 & 0.19\\
\midrule\multicolumn{8}{l}{\emph{Technical implications (baseline: none)}}\\
 Reduced battery life & \includegraphics[page=5]{figures/tab_clmm_results_vignettes.pdf} & \textbf{-0.15} & \textbf{< 0.01} & \textbf{-0.14} & \textbf{< 0.01} & -0.04 & 0.19\\
 Malfunction contact tracing & \includegraphics[page=6]{figures/tab_clmm_results_vignettes.pdf} & \textbf{-0.19} & \textbf{< 0.01} & \textbf{-0.20} & \textbf{< 0.01} & -0.09 & 0.19\\
 Malfunction information & \includegraphics[page=7]{figures/tab_clmm_results_vignettes.pdf} & -0.03 & 0.64 & -0.12 & 0.07 & -0.09 & 0.20\\
 Malfunction quarantine enforcement & \includegraphics[page=8]{figures/tab_clmm_results_vignettes.pdf} & \textbf{-0.20} & \textbf{< 0.01} & \textbf{-0.25} & \textbf{< 0.01} & \textbf{-0.24} & \textbf{< 0.01}\\
 Malfunction symptom check & \includegraphics[page=9]{figures/tab_clmm_results_vignettes.pdf} & -0.04 & 0.56 & -0.09 & 0.18 & -0.13 & 0.05\\
 Malfunction health certificate & \includegraphics[page=10]{figures/tab_clmm_results_vignettes.pdf} & \textbf{-0.16} & \textbf{0.01} & \textbf{-0.32} & \textbf{< 0.01} & \textbf{-0.15} & \textbf{0.02}\\
\midrule\multicolumn{8}{l}{\emph{Societal implications (baseline: none)}}\\
 Personal advantages & \includegraphics[page=11]{figures/tab_clmm_results_vignettes.pdf} &  0.00 & 0.97 & -0.05 & 0.40 &  & \\
 Faster update of regional rules & \includegraphics[page=12]{figures/tab_clmm_results_vignettes.pdf} & \textbf{0.14} & \textbf{< 0.01} & \textbf{0.09} & \textbf{0.03} &  & \\
 Future use cases & \includegraphics[page=13]{figures/tab_clmm_results_vignettes.pdf} & -0.00 & 0.95 & -0.01 & 0.72 &  & \\
\midrule\multicolumn{8}{l}{\emph{Payload data (baseline: none)}}\\
 Encounters & \includegraphics[page=14]{figures/tab_clmm_results_vignettes.pdf} & \textbf{0.06} & \textbf{0.04} &  &  &  & \\
 Location & \includegraphics[page=15]{figures/tab_clmm_results_vignettes.pdf} & -0.04 & 0.16 &  &  & 0.04 & 0.22\\
 Infection Status & \includegraphics[page=16]{figures/tab_clmm_results_vignettes.pdf} & 0.04 & 0.22 & 0.05 & 0.21 &  & \\
 Health Information & \includegraphics[page=17]{figures/tab_clmm_results_vignettes.pdf} &  &  & 0.03 & 0.33 & -0.04 & 0.18\\
 Unspecified & \includegraphics[page=18]{figures/tab_clmm_results_vignettes.pdf} &  &  &  &  & \textbf{-0.34} & \textbf{0.01}\\
\midrule\multicolumn{8}{l}{\emph{Identification data (baseline: none)}}\\
 Demographic data & \includegraphics[page=19]{figures/tab_clmm_results_vignettes.pdf} & \textbf{-0.28} & \textbf{< 0.01} & \textbf{-0.20} & \textbf{< 0.01} & 0.09 & 0.05\\
 Unique identification of individual & \includegraphics[page=20]{figures/tab_clmm_results_vignettes.pdf} & \textbf{-0.30} & \textbf{< 0.01} & \textbf{-0.16} & \textbf{< 0.01} & \textbf{0.10} & \textbf{0.02}\\
\midrule\multicolumn{8}{l}{\emph{Data receiver (baseline: unspecified)}}\\
 Health authorities & \includegraphics[page=21]{figures/tab_clmm_results_vignettes.pdf} & -0.02 & 0.70 & -0.00 & 0.92 & 0.08 & 0.08\\
 Research institutes & \includegraphics[page=22]{figures/tab_clmm_results_vignettes.pdf} & -0.02 & 0.66 & 0.03 & 0.48 & 0.01 & 0.76\\
 Private companies & \includegraphics[page=23]{figures/tab_clmm_results_vignettes.pdf} & \textbf{-0.29} & \textbf{< 0.01} & \textbf{-0.17} & \textbf{< 0.01} & \textbf{-0.17} & \textbf{< 0.01}\\
 Law enforcement & \includegraphics[page=24]{figures/tab_clmm_results_vignettes.pdf} & \textbf{-0.32} & \textbf{< 0.01} & \textbf{-0.26} & \textbf{< 0.01} & 0.06 & 0.21\\
 Public & \includegraphics[page=25]{figures/tab_clmm_results_vignettes.pdf} & \textbf{-0.30} & \textbf{< 0.01} & -0.05 & 0.45 & -0.07 & 0.30\\
\midrule\multicolumn{8}{l}{\emph{Data transmission (baseline: automatically)}}\\
 Manual & \includegraphics[page=27]{figures/tab_clmm_results_vignettes.pdf} & 0.04 & 0.15 & \textbf{0.08} & \textbf{< 0.01} &  & \\
\midrule\multicolumn{8}{l}{\emph{Retention period (baseline: unspecified)}}\\
 One month & \includegraphics[page=28]{figures/tab_clmm_results_vignettes.pdf} &  &  & 0.06 & 0.09 & \textbf{0.08} & \textbf{0.02}\\
 End of COVID-19 pandemic & \includegraphics[page=29]{figures/tab_clmm_results_vignettes.pdf} &  &  & \textbf{0.09} & \textbf{0.01} & 0.03 & 0.35\\

\bottomrule
\end{tabular*}
\end{table*}
\begin{table*}[htbp]
\Description[Non-scenario-based CLMM regression table part 1 of participants' willingness to use corona apps]{
Non-scenario-based CLMM regression table part 1 has the same structure as the scenario-based CLMM regression table.
It consists of eight columns: a column to label the factor and the factor level in each row, a column depicting the effect size as bar chart (one bar for each country), and six data columns---two for each country. The data columns form three pairs of an estimate and a p-value column. The table header labels reads as ``Factor'', ``+/-'' (increase or decrease), and the country names where each country has a sub header ``Estimate'' and ``Pr(>|z|)''. The rows are grouped by eight factors and the first row in each of these groups contains only the factor name and the baseline factor level. The other rows contain the factor levels and the corresponding data.
}
\centering
\caption[CLMM regression of participants' willingness to use corona apps. (Full)]{\label{tab:clmm_results}
Cumulative link mixed model regression for participants' willingness to use corona apps. A positive estimate (effect size) indicates the willingness to use being higher compared to the factor's baseline variable. All factors were collected as responses in the non-scenario parts of the questionnaire. Only factors present in the final model are presented here The colored bars represent the effect sizes per country (yellow: Germany, blue: United States, red: China).}
\begin{tabular*}{\textwidth}{@{}ll@{\extracolsep{\fill}}*{6}{r}@{}}
\toprule
Factors and Levels & \multicolumn{1}{c}{+/-} & \multicolumn{2}{c}{\textbf{Germany}} & \multicolumn{2}{c}{\textbf{United States}} & \multicolumn{2}{c}{\textbf{China}} \\
&& Estimate & Pr(>|z|) & Estimate & Pr(>|z|) & Estimate & Pr(>|z|) \\

\midrule\multicolumn{8}{@{}l@{}}{\emph{[Q2]: Phone OS (baseline: Android)}}\\
 iOS & \includegraphics[page=1]{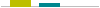} & 0.25 & 0.07 & 0.13 & 0.30 &  & \\
\midrule\multicolumn{8}{@{}l@{}}{\emph{[Q3r1]: Satisfied with battery life (baseline: neutral) }}\\
 satisfied & \includegraphics[page=2]{figures/tab_clmm_results.pdf} &  &  & -0.25 & 0.19 & 0.13 & 0.41\\
 dissatisfied & \includegraphics[page=3]{figures/tab_clmm_results.pdf} &  &  & \textbf{-0.57} & \textbf{0.02} & \textbf{-0.61} & \textbf{< 0.01}\\
\midrule\multicolumn{8}{@{}l@{}}{\emph{[Q3r2]: Satisfied with location accuracy (baseline: neutral) }}\\
 satisfied & \includegraphics[page=4]{figures/tab_clmm_results.pdf} & \textbf{0.44} & \textbf{< 0.01} & \textbf{0.54} & \textbf{< 0.01} &  & \\
 dissatisfied & \includegraphics[page=5]{figures/tab_clmm_results.pdf} & \textbf{1.21} & \textbf{0.01} & 0.64 & 0.23 &  & \\
\midrule\multicolumn{8}{@{}l@{}}{\emph{[Q4]: Infected participant (baseline: not tested, no infection suspected)}}\\
 Tested negative only & \includegraphics[page=6]{figures/tab_clmm_results.pdf} &  &  & 0.21 & 0.26 &  & \\
 Not tested, infection suspected & \includegraphics[page=7]{figures/tab_clmm_results.pdf} &  &  & \textbf{0.58} & \textbf{0.02} &  & \\
 Tested positive & \includegraphics[page=8]{figures/tab_clmm_results.pdf} &  &  & 0.77 & 0.09 &  & \\
\midrule\multicolumn{8}{@{}l@{}}{\emph{[Q5]: Infection in social circles (baseline: no)}}\\
 Yes & \includegraphics[page=10]{figures/tab_clmm_results.pdf} &  &  & -0.32 & 0.09 &  & \\
\midrule\multicolumn{8}{@{}l@{}}{\emph{[Q6]: Quarantine experience (baseline: no)}}\\
 Yes & \includegraphics[page=11]{figures/tab_clmm_results.pdf} &  &  &  &  & -0.21 & 0.11\\
\midrule\multicolumn{8}{@{}l@{}}{\emph{[Q7]: Risk person in household (baseline: no)}}\\
 Yes & \includegraphics[page=12]{figures/tab_clmm_results.pdf} &  &  &  &  & \textbf{0.24} & \textbf{0.04}\\
\midrule\multicolumn{8}{@{}l@{}}{\emph{[Q8]: Infection concerns (baseline: somewhat concerned)}}\\
 Slightly or not concerned & \includegraphics[page=13]{figures/tab_clmm_results.pdf} & \textbf{-0.62} & \textbf{< 0.01} & \textbf{-0.55} & \textbf{< 0.01} & -0.08 & 0.57\\
 Moderately or exteremly concerned & \includegraphics[page=14]{figures/tab_clmm_results.pdf} & 0.26 & 0.05 & \textbf{0.54} & \textbf{< 0.01} & 0.19 & 0.17\\
\midrule\multicolumn{8}{@{}l@{}}{\emph{[Q13r2]: Knows app for symptom check  (baseline: yes)}}\\
 No & \includegraphics[page=15]{figures/tab_clmm_results.pdf} & \textbf{-0.36} & \textbf{0.04} & -0.37 & 0.09 & \textbf{-0.35} & \textbf{0.01}\\
\midrule\multicolumn{8}{@{}l@{}}{\emph{[Q13r3]: Knows app for quarantine enforcement (baseline: yes)}}\\
 No & \includegraphics[page=17]{figures/tab_clmm_results.pdf} & -0.42 & 0.05 &  &  & \textbf{-0.42} & \textbf{< 0.01}\\
\midrule\multicolumn{8}{@{}l@{}}{\emph{[Q13r4]: Knows app for contact tracing (baseline: yes)}}\\
 No & \includegraphics[page=19]{figures/tab_clmm_results.pdf} & \textbf{0.44} & \textbf{0.01} & -0.46 & 0.06 & \textbf{-0.57} & \textbf{< 0.01}\\
\midrule\multicolumn{8}{@{}l@{}}{\emph{[Q13r5]: Knows app for health certificate (baseline: yes)}}\\
 No & \includegraphics[page=21]{figures/tab_clmm_results.pdf} & \textbf{-0.66} & \textbf{< 0.01} & \textbf{-0.80} & \textbf{< 0.01} &  & \\

\bottomrule
\end{tabular*}
\end{table*}

\begin{table*}[htbp]
\Description[Non-scenario-based CLMM regression table part 2 of participants' willingness to use corona apps]{
Non-scenario-based CLMM regression table part 2 has the same structure as non-scenario-based CLMM regression table part 1.
It consists of eight columns: a column to label the factor and the factor level in each row, a column depicting the effect size as bar chart (one bar for each country), and six data columns---two for each country. The data columns form three pairs of an estimate and a p-value column. The table header labels reads as ``Factor'', ``+/-'' (increase or decrease), and the country names where each country has a sub header ``Estimate'' and ``Pr(>|z|)''. The rows are grouped by eight factors and the first row in each of these groups contains only the factor name and the baseline factor level. The other rows contain the factor levels and the corresponding data.
}
\centering
\caption[CLMM regression of participants' willingness to use corona apps. (Full --Continued)]{\label{tab:clmm_results2}
Cumulative link mixed model regression for participants' willingness to use corona apps (continued from Table~\ref{tab:clmm_results}).}
\begin{tabular*}{\textwidth}{@{}ll@{\extracolsep{\fill}}*{6}{r}@{}}
\toprule
Factors and Levels & \multicolumn{1}{c}{+/-} & \multicolumn{2}{c}{\textbf{Germany}} & \multicolumn{2}{c}{\textbf{United States}} & \multicolumn{2}{c}{\textbf{China}} \\
&& Estimate & Pr(>|z|) & Estimate & Pr(>|z|) & Estimate & Pr(>|z|) \\

\midrule\multicolumn{8}{@{}l@{}}{\emph{[Q14]: Uses any corona app (baseline: yes)}}\\
 No & \includegraphics[page=1]{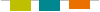} & \textbf{-0.83} & \textbf{< 0.01} & \textbf{-0.99} & \textbf{< 0.01} & \textbf{-0.30} & \textbf{0.02}\\
\midrule\multicolumn{8}{@{}l@{}}{\emph{[Q17]: Rate measures (baseline: too strict)}}\\
 About right                            & \includegraphics[page=4]{figures/tab_clmm_results2.pdf} & 0.24 & 0.12 &  &  &  & \\
 Too lenient                            & \includegraphics[page=5]{figures/tab_clmm_results2.pdf} & \textbf{0.71} & \textbf{< 0.01} &  &  &  & \\
\midrule\multicolumn{8}{@{}l@{}}{\emph{[Q18r1]: Opinion of health authorities (baseline: neutral)}}\\
 Favorable & \includegraphics[page=6]{figures/tab_clmm_results2.pdf} & \textbf{0.40} & \textbf{< 0.01} &  &  & 0.13 & 0.50\\
 Unfavorable & \includegraphics[page=7]{figures/tab_clmm_results2.pdf} & 0.02 & 0.94 &  &  & 0.21 & 0.56\\
\midrule\multicolumn{8}{@{}l@{}}{\emph{[Q18r2]: Opinion of law enforcement (baseline: neutral)}}\\
 Favorable & \includegraphics[page=8]{figures/tab_clmm_results2.pdf} &  &  & 0.15 & 0.31 & 0.07 & 0.70\\
 Unfavorable & \includegraphics[page=9]{figures/tab_clmm_results2.pdf} &  &  & \textbf{0.44} & \textbf{0.03} & 0.50 & 0.07\\
\midrule\multicolumn{8}{@{}l@{}}{\emph{[Q18r3]: Opinion of research institutions (baseline: neutral)}}\\
 Favorable & \includegraphics[page=10]{figures/tab_clmm_results2.pdf} &  &  & \textbf{0.44} & \textbf{0.01} & 0.50 & 0.07\\
 Unfavorable & \includegraphics[page=11]{figures/tab_clmm_results2.pdf} &  &  & -0.10 & 0.66 & 0.50 & 0.07\\
\midrule\multicolumn{8}{@{}l@{}}{\emph{[Q18r4]: Opinion of private companies (baseline: neutral)}}\\
 Favorable & \includegraphics[page=12]{figures/tab_clmm_results2.pdf} & 0.22 & 0.08 &  &  & \textbf{0.27} & \textbf{0.03}\\
 Unfavorable & \includegraphics[page=13]{figures/tab_clmm_results2.pdf} & -0.13 & 0.43 &  &  & -0.22 & 0.33\\
\midrule\multicolumn{8}{@{}l@{}}{\emph{[Q18r5]: Opinion of federal government (baseline: neutral)}}\\
 Favorable & \includegraphics[page=14]{figures/tab_clmm_results2.pdf} & -0.20 & 0.27 & 0.35 & 0.04 &  & \\
 Unfavorable & \includegraphics[page=15]{figures/tab_clmm_results2.pdf} & \textbf{-0.49} & \textbf{0.02} & 0.15 & 0.43 &  & \\
\midrule\multicolumn{8}{@{}l@{}}{\emph{[Q18r6]: Opinion of state government (baseline: neutral)}}\\
 Favorable & \includegraphics[page=16]{figures/tab_clmm_results2.pdf} & \textbf{0.49} & \textbf{< 0.01} & \textbf{0.35} & \textbf{0.04} &  & \\
 Unfavorable & \includegraphics[page=17]{figures/tab_clmm_results2.pdf} & 0.20 & 0.31 & 0.16 & 0.40 &  & \\
\midrule\multicolumn{8}{@{}l@{}}{\emph{[Q19]: Acceptability of sharing network data  (baseline: neutral)}}\\
 Acceptable & \includegraphics[page=18]{figures/tab_clmm_results2.pdf} & \textbf{0.57} & \textbf{< 0.01} & \textbf{0.72} & \textbf{< 0.01} & \textbf{0.36} & \textbf{0.01}\\
 Unacceptable & \includegraphics[page=19]{figures/tab_clmm_results2.pdf} & \textbf{-0.59} & \textbf{< 0.01} & \textbf{-0.76} & \textbf{< 0.01} & -0.07 & 0.76\\
  \midrule\multicolumn{8}{@{}l@{}}{\emph{[Q20]: IUIPC}}\\
Control & \includegraphics[page=20]{figures/tab_clmm_results2.pdf} &  &  & \textbf{-0.11} & \textbf{0.05} & \textbf{0.46} & \textbf{< 0.01}\\
 Awareness & \includegraphics[page=21]{figures/tab_clmm_results2.pdf} &  &  &  &  & \textbf{0.18} & \textbf{0.02}\\
 Collection & \includegraphics[page=22]{figures/tab_clmm_results2.pdf} & \textbf{-0.16} & \textbf{< 0.01} &  &  & -0.10 & 0.10\\
\midrule\multicolumn{8}{@{}l@{}}{\emph{Gender (baseline: Female)}}\\
 Male & \includegraphics[page=23]{figures/tab_clmm_results2.pdf} & \textbf{0.23} & \textbf{0.03} & \textbf{0.26} & \textbf{0.05} &  & \\
\midrule Age & \includegraphics[page=24]{figures/tab_clmm_results2.pdf} &  &  & \textbf{-0.02} & \textbf{< 0.01} &  & \\
\midrule\multicolumn{8}{@{}l@{}}{\emph{Education (baseline: Less than high school)}}\\
 High school or associate degree & \includegraphics[page=25]{figures/tab_clmm_results2.pdf} &  &  & -0.33 & 0.19 & 0.17 & 0.65\\
 Bachelor's degree  & \includegraphics[page=26]{figures/tab_clmm_results2.pdf} &  &  & -0.49 & 0.05 & -0.24 & 0.51\\
 Postgraduate Education & \includegraphics[page=27]{figures/tab_clmm_results2.pdf} &  &  & -0.37 & 0.23 & 0.02 & 0.96\\

\bottomrule
\end{tabular*}
\end{table*}

\subsubsection{Germany}

In Germany, being male was the only demographic factor with a significant positive influence on participants' willingness to use a corona app, while education level and age were excluded from the model for not helping explain the decision to use an app. 

Higher privacy concern with regard to data collection practices (IUIPC: Collection) had a significant negative influence on the decision to use an app.

Strong opinions on the location accuracy of the phone~(Q3), both positive and negative, had a significant positive influence on the willingness to use any of the apps.
Previous use of corona apps and knowing apps for symptom checks or health certificates had a significant positive impact on the willingness to use a corona app~(Q13).
Conversely, knowing apps for contact tracing had a significant negative impact~(Q14).
Additionally, rating the practice of companies sharing customers' data with authorities to help fight the pandemic as acceptable ~(Q19) had a significant positive impact on the willingness to use an app in any scenario.
The level of concern about becoming infected with coronavirus~(Q8) had a significant impact, with less concerned participants being less likely to use any app. 
Rating the measures taken to fight the COVID-19 pandemic as too lenient~(Q17) had a significant positive impact on the willingness to use an app. The results further suggest that a favorable rating of institutions (health authorities and state government) had a significant and positive impact on the decision to use any of the apps, whereas rating them unfavorably (federal government) had a negative impact~(Q18). 

Within the app scenarios, contact tracing was the only app purpose with a significant positive influence on participants' willingness to use a corona app. Technical implications often negatively impacted participants' willingness to use the described app, and did so significantly in the case of increased battery consumption as well as malfunctions in contact tracing, quarantine enforcement, and health certificate apps. This supports our descriptive results of participants being least satisfied with their smartphones' battery life. 
Scenarios with the societal implication of a more timely adjustment of local coronavirus regulations also had a significant positive impact on the willingness to use an app. 
Regarding the specific types of data used by the fictitious apps, encounter data significantly positively influenced participants' decisions to use an app while the collection of demographic data and data that could be used to uniquely identify the user was opposed.
German participants also had a strong opinion about who should be the data receiver. When scenarios described data transfers to private companies, law enforcement, or the general public, participants were significantly less likely to use the app.

\subsubsection{United States}

In the United States, male gender had a significant positive influence on the willingness to use any of the presented apps, as did higher age, although to a lesser degree. Similar to the German data, participants' education level had little to no impact. 

Participants with higher privacy concern regarding control of their personal information (IUIPC: Control) were significantly less likely to use an app.  
In contrast to the model for the German data set, the suspected coronavirus infection had a significant positive impact on the decision to use any of the apps~(Q4) in the US.
As for the German sample, rating as acceptable the practice of companies sharing customers' data with authorities to help fight the pandemic~(Q19) had a significant positive impact on the willingness to use an app. Similarly, the results show that favorably rating some institutions (research institutions and state governments) had a significant and positive impact on app adoption, while, interestingly, unfavorable ratings of law enforcement also had a significant positive impact on the willingness to use an app~(Q18).
Concern about becoming infected with coronavirus~(Q8) was a significant influence on the decision to use any of the presented apps: Higher levels of concern about an infection positively influenced the willingness to use an app while lower levels had a significant negative impact. Similar to the German model, prior use of any corona app, as well as knowledge of health certificate apps, had a significantly positive  impact on the willingness to use the apps~(Q13,~14).  
Regarding the attitude towards certain properties of one's phone, satisfaction with the phone's location accuracy and dissatisfaction with its battery life had significant positive and negative impacts, respectively, on US participants' willingness to use an app, the former being similar to the German sample.

Looking at the influence of the factor levels in the app scenarios, we found the app purpose of contact tracing to have a significant positive influence on US participants' willingness to use an app, as in the German sample. However, in the US a positive influence, to a lesser degree but still significant, was also found for the app purposes of symptom checks and health certificates.  
When scenarios described manual data transmission, participants were slightly but significantly more willing to use the app compared to scenarios with automatic data transmission. The societal implication of a more timely adjustment of local coronavirus regulations also had a significant positive influence on the willingness to use the app, while the other societal implications did not have any impact. Compared to scenarios that did not mention the retention period, if an app's data storage was limited to the end of the coronavirus pandemic, participants were significantly more willing to use it. 
Similar to the German model, malfunctions in contact tracing, quarantine enforcement, and health certificate apps, as well as increased battery consumption, had a significant negative impact on the willingness to use a corona app. US participants had a similar opinion as Germans about who should be the data receiver: Scenarios with data being sent to private companies or law enforcement significantly negatively influenced the willingness to use an app compared to scenarios that did not mention who would receive the data. 
In contrast to the model for the German data set, specific types of payload data did not significantly impact participants' willingness to use an app, while participants from both countries share the negative attitude towards corona apps that collect demographic data or data that allows for the unique identification of the individual.

\subsubsection{China}

In China, demographic information was no good predictor for participants' willingness to use any type of corona app: Age and gender were not part of the final model, and similar to the US model, education had no significant impact.

Participants with higher privacy concern (IUIPC: Control and Awareness) were found to be significantly more likely to use corona apps. An individual's infection status was not part of the final model for the Chinese data.
Individual infection concerns (Q8) did not have any significant impact on app use. However, living with a person at higher risk (Q7) significantly positively influenced the willingness to use an app.
Similar to the other models, prior usage of any corona app (Q14), as well as knowledge of apps for symptom checks, quarantine enforcement, and contact tracing (Q13), positively influenced the willingness to use an app.
As in the US, dissatisfaction with the battery life of one's phone (Q3) had a significant negative impact on the willingness to use any of the presented apps. 
Additionally, as for the German and the US samples, rating as acceptable that companies share customers' data with authorities to help fight the pandemic (Q19) had a significant positive impact on the willingness to use the app in any of the scenarios, as did a favorable opinion of private companies (Q18). 

Scenario-based factors were also influential: In the Chinese model, the app purposes of quarantine enforcement and symptom check had a significant negative influence on the willingness to use an app. The effect size for quarantine enforcement was larger than in the other countries, where it was also not significant. Regarding the types of payload data only scenarios where the data type was not specified had a significant negative influence on the decision to use the app. In stark contrast to the results from the US and Germany, Chinese participants were significantly more willing to use the presented app if it collected data that could be used to uniquely identify the person. Regarding potential receivers of the data we computed a significant negative effect size for private companies, and participants slightly favored shorter retention times.
Similar to the results of the models for the German and the US samples, malfunctions of quarantine enforcement and health certificate apps significantly negatively influenced Chinese participants' willingness to use them. 

\begin{table*}[htbp]
\Description[Descriptive statistic table of reasons not to use a corona app]{
The reasons not to use a corona app table consists of nine columns: a theme label column, a code label column, an examples column, three ``WhyNoApp'' columns (one per country), and three ``Negative'' columns (one per country). A theme compromise several codes, i.e., a theme label table cell may span multiple rows. The first two rows are header rows. The next two rows contains the frequency of non-app users and non-empty answers (and have therefore no theme label). The other rows are grouped by seven themes and have a code label, one to three example texts for the code, and the frequencies of answers label with this code per country and the categories ``WhyNoApp'' and ``Negative''.
}
\centering
\caption[Reported reasons not to use a corona app and negative aspects.]{\label{tab:quali-negative}
Reported reasons not to use a corona app (Q14a) and negative aspects of corona apps (Q16). Q14a was only shown to participants who had indicated not to use a corona app in Q14 (``non-users of app'').
}

\setlength{\tabcolsep}{2pt}
\begin{tabular*}{\textwidth}{@{}>{\bfseries}cll@{\extracolsep{\fill}}*{6}{r}}
\toprule
& \multirow{2}{*}{\textbf{Code}} & \multirow{2}{*}{\textbf{Examples}} & \multicolumn{3}{@{}c@{}}{\textbf{Q14a}} & \multicolumn{3}{@{}c@{}}{\textbf{Q16}} \\
&&& \textbf{DE} & \textbf{US} & \textbf{CN} & \textbf{DE} & \textbf{US} & \textbf{CN} \\
\midrule 
& Non-users of app & & 961 & 937 & 405 & -- & -- & -- \\
& Non-empty answers & & 685 & 702 & 276 & 726 & 760 & 738 \\
\midrule
\addlinespace[.5\defaultaddspace]
\multirow{4}{*}{\rotatebox[origin=c]{90}{\parbox[c]{35pt}{\centering Privacy~/ Se\-cu\-ri\-ty}}}
& (govt.) surveillance & ``Big Brother'', ``used for more than just corona''                             & 53              & 20         & 1                 & 174            & 70             & 1        \\
\addlinespace[.75\defaultaddspace]
& privacy       & ``leak information'', ``invasive''                        & 72              & 96         & 12                 & 292            & 337             & 179        \\
\addlinespace[.75\defaultaddspace]
& security      & ``too insecure'', ``could be hacked''                    & 21              & 5         & 1                 & 18            & 15             & 1        \\
\addlinespace[.75\defaultaddspace]
\midrule\multirow{4}{*}{\rotatebox[origin=c]{90}{\parbox{30pt}{\centering Avail\-abil\-i\-ty}}}
& no app            & ``didn't know there's such an app'', ``no app yet''                & 214              & 222         & 73                 & --            & --             & --        \\
& no suitable app          & ``don't know a good one'', ``no approved [...] app''                 & 61              & 18         & 13                 & --            & --             & --        \\
& lack of information      & ``don't know how to use it'', ``not familiar''                  & 31              & 27         & 19                 & --            & --             & --        \\
& no phone         & ``I do not have a smartphone''        & 5              & 13         & 0                 & --            & --              &  --        \\

\midrule\multirow{7}{*}{\rotatebox[origin=c]{90}{\parbox{60pt}{\centering Psychological~/ Societal}}}
& user base         & ``not enough users'', ``need to be used correctly''                   & 11              & 4         & 4                 & 48            & 54             & 28        \\
& discrimination        & ``division into good and evil", ``stigmatization''                    & 2              & 2         & 0                 & 9            & 9             & 2        \\
& disinformation         & ``more fake news'', ``government may hide the truth''                   & 0              & 0\         & 0                 & 1            & 7             & 39        \\
& anxiety           & ``would make me more nervous'', ``freaks people out''                 & 8              & 6         & 4                 & 12            & 20              &  69        \\
& false sense of protection     & ``blind faith in the [...] app''                       & 0              & 0         & 0                 & 4            & 4              &  1        \\
& autonomy         & ``loss of freedom'', ``civil rights'', ``unconstitutional''                   & 7              & 14         & 0                 & 36            & 38              &  5        \\
& not mandatory        & ``because I don't need to install it'', ``not compulsory''                    & 0              & 1         & 5                 & 0            & 0              &  3        \\

\midrule\multirow{5}{*}{\rotatebox[origin=c]{90}{Technical}}
& phone use         & ``do not carry phone all the time'', ``no mobile data''                   & 21              & 29         & 1                 & --            & --             & --        \\
& not supported         & ``my phone is too old'', ``does not support new apps''             & 5              & 5         & 0                 & --            & --             & --        \\
& technical side effects            & ``drains battery too much'', ``use[s] up the memory''                & 9              & 22         & 4                 & 21            & 37             & 12        \\
& malfunctions            & ``not reliable'', ``not [...] accurate'', ``false positives''                & 12              & 12         & 5                 & 32            & 64              &  48        \\
& inconvenient          & ``too complicated'', ``need to sign in every day''                  & 3              & 2         & 7                 & 2            & 1              &  13        \\

\midrule\multirow{7}{*}{\rotatebox[origin=c]{90}{Unnecessary}}
& state of the pandemic           & ``too late'', ``low risk region''                 & 8              & 5         & 18                 & 2            & 0             & 1        \\
& personal behavior             & ``I rarely go out'', ``my range of activities is small''               & 11              & 34         & 13                 & --            & --             & --        \\
& plugin           & ``there [are] similar functions in WeChat''                 & 0              & 0         & 32                 & 0            & 0             & 1        \\
& other information sources       & ``I watch TV'', ``I go to websites for information''                     & 15              & 24         & 3                 & 1            & 0              &  0        \\
& other measures          & ``social distancing'', ``I wear a mask''                  & 14              & 32         & 4                 & 2            & 2              &  0        \\
& coronaskeptic        & ``corona is fake'', ``just a flu'', ``conspiracy''                   & 12              & 16         & 0                 & 7            & 5              &  0        \\
& unnecessary (general)        & ``useless'', ``a waste of time'', ``I don't need it''                    & 96              & 103         & 56                 & 9            & 18              &  12        \\

\midrule\multirow{4}{*}{\rotatebox[origin=c]{90}{Other}}
& don't want            & ``I just don't want to'', ``no'', ``stupid''                & 23              & 36         & 5                 & --            & --             & --        \\
& generic negative            & ``everything'', ``yes'', ``many''                & --              & --         & --                 & 6            & 21             & 4        \\
& other          & ``too much information'', ``I do not trust them''                 & 14              & 20         & 3                 & 17            & 24             & 25        \\
& none          & ``I don't see any'', ``nothing really''          & --              & --         & --                 & 87            & 72             & 232        \\

\midrule\multirow{4}{*}{\rotatebox[origin=c]{90}{\parbox{40pt}{\centering Non-answers}}}
& don't know          & ``don't know'', ``not sure'', ``hard to tell''                  & 10              & 12         & 8                 & 20            & 33             & 28        \\
& positive           & ``help[s] keep people safe'', ``practical'', ``very good''                 & --              & --         & --                 & 5            & 14              &  45        \\
& unclear           & ``data problem'', ``Facebook'', ``more abundant''                 & 13              & 16         & 10                 & 18            & 22              &  32        \\
& no answer          & ``./.'', ``-'', ``vwedv'', ``Don't answer at the present time''                  & 8              & 2         & 0                 & 11            & 19              &  3        \\

\bottomrule
\end{tabular*}
\end{table*}
\begin{table*}[htbp]
\Description[Descriptive statistic table of reported positive aspects of corona apps]{
The reported positive aspects of corona apps table consists of six columns: a theme label column, a code label column, an examples column, three frequency columns (one per country). A theme compromise several codes, i.e., a theme label table cell may span multiple rows The table header labels the second column with ``Code'', the third column with ``Examples'', and the frequency columns with Germany, US, and China. The frequency column have also a sub header showing the total number of answers to Q15 per country. The rows are grouped by four themes and have a code label, one to three example texts for the code, and the frequencies of answers label with this code per country.
}
\centering
\caption[Reported positive aspects of corona apps (Q15)]{\label{tab:quali-positive}
Reported positive aspects of corona apps (Q15)
}

\begin{tabular*}{\textwidth}{@{}>{\bfseries}cll@{\extracolsep{\fill}}*{3}{r}}
\toprule
&  \textbf{Code} & \textbf{Examples} & \textbf{DE} & \textbf{US} & \textbf{CN} \\
\midrule
& Non-empty answers & & 738 & 759 & 780 \\
\midrule\multirow{6}{*}{\rotatebox[origin=c]{90}{App Purposes}}

& contact tracing      & ``to trace affected people''                              & 106  & 103  &  19 \\ 
& symptom check        & ``help identify symptoms''                                &   8  &  22  &  13 \\ 
& quarantine           & ``enforcing quarantine''                                  &   5  &   9  &  10 \\ 
& information          & ``know the situation around''                             &  60  &  91  & 213 \\ 
& health certificate   & ``health QR code'', ``you can show your status''          &   0  &   1  &  40 \\ 
& movement             & ``record movement tracks'', ``travel route''              &   2  &  12  &  87 \\ 

\midrule
\rotatebox[origin=c]{90}{\parbox{30pt}{\centering Avail\-a\-bil\-i\-ty}}
& availability         & ``everyone has a smartphone'', ``convenient''             &  27  &  31  &  22 \\ 
\midrule
\addlinespace[1.25\defaultaddspace]
\multirow{2}{*}{\rotatebox[origin=c]{90}{\parbox{30pt}{\centering Psycho\-logical}}}
& sense of protection  & ``makes one feel safer'', ``relieve stress''              &   8  &   3  &  12 \\ 
\addlinespace[1.25\defaultaddspace]
& awareness            & ``remind us'', ``take it more seriously''                 &   5  &  15  &  35 \\ 
\addlinespace[1.25\defaultaddspace]

\midrule
\addlinespace[.25\defaultaddspace]
\multirow{4.5}{*}{\rotatebox[origin=c]{90}{\parbox{40pt}{\centering Pandemic Control}}}
& measures             & ``assess the risk levels [...] and act accordingly''      &  28  &   4  &  13 \\ 
\addlinespace[.25\defaultaddspace]
& infection detection  & ``knowledge of hot spots'', ``identify infected people''  &  42  &  72  &  84 \\ 
\addlinespace[.25\defaultaddspace]
& infection prevention & ``reduce the risk of infection'', ``keep people safe''    &  41  &  19  &  70 \\ 
\addlinespace[.25\defaultaddspace]
& generic pandemic     &  ``limit the spread'', ``control the outbreak''           &  74  &  61  &  55 \\ 
\addlinespace[.25\defaultaddspace]

\midrule
\multirow{3}{*}{\rotatebox[origin=c]{90}{Other}}
& generic positive     & ``very good'', ``it helps'', ``many''                     &  57  &  56  & 100 \\ 
& other                & ``big data'', ``publicity'', ``no more home office''      &  18  &  10  &  35 \\ 
& none                 & ``none'', ``not really'', ``nothing''                     & 212  & 170  &  16 \\ 

\midrule
\multirow{4}{*}{\rotatebox[origin=c]{90}{\parbox{35pt}{\centering Non-answers}}}
& don't know           & ``don't know'', ``hard to say'', ``not sure''             &  27  &  64  &  13 \\ 
& negative             & ``virus gets on phone'', ``invasion of privacy''          &  63  &  69  &  17 \\ 
& unclear              & ``Ues it impacts'', ``Japanese style'', ``ccorona''       &  25  &  32  &  49 \\ 
& no answer            & ``jhgfkjfkuz'', ``Tv'', ``na'', ``??''                    &  14  &  17  &   4 \\ 

\bottomrule
\end{tabular*}
\end{table*}
\subsection{Perception of Corona Apps} 

\subsubsection{Use of Corona Apps}


As depicted in Table~\ref{tab:demographics}, the number of participants who reported to already use a corona app (DE: \num{42} / \SI{4.2}{\percent}, US: \num{66} / \SI{6.6}{\percent}, CN: \num{614} / \SI{60.3}{\percent}) reflects whether government-recommended apps already existed at the time of the study. Those who reported to use an app were asked in Q14b which app they used. 
In Germany, 25 participants provided insightful answers. Two mentioned using a specific app but could not remember the name, one mentioned an app for general news, five a non-app news source, another five Germany's official disaster information app, NINA, that has been extended to provide coronavirus-related news (see Section~\ref{subsubsec:germany}), one a non-corona-specific health app, and 11 named coronavirus-related apps, 9 of which we could map to ``Corona-Datenspende'', the data donation app launched by German health authorities~\cite{rki_coronadatenspende_2020}.
In the US, we received 21 answers. Seven participants named COVID-19-specific apps (Apple's COVID-19 app, C Spire Health, Healthy Together, and How We Feel), two disaster information apps, four general health apps, another four a non-health-related app or service (such as ``Facebook''), and two non-app news sources.
In China, \num{153} participants reported to use health QR code apps issued by various local or regional authorities, which are often WeChat or Alipay plugins, and \num{108} specifically stated using (plugins for) WeChat or Alipay. 
\num{15} participants mentioned to use other platform economies, \eg, Weibo or Tencent, in a COVID-19-related context.
An additional \num{49} participants named further COVID-specific apps, and \num{27} reported using non-corona-specific health apps.


\subsubsection{Reasons not to Use Corona Apps / Perceived Negative Aspects of Corona Apps}

During the coding process of the open-ended answers, we observed a large overlap between the reasons not to use a corona app~(Q14a) and perceived negative aspects of such apps~(Q16). This led us to devise a common coding frame for both questions. 
Our combined codebook for Q14a and Q16 contained 34 codes.
Inter-coder reliability as determined by Krippendorff's alpha was in the $ (0.66, 1) $ range for individual codes, with the rather generic ``unclear'' code scoring lowest and a weighted mean of $ 0.91 $. 
Q14a was only displayed to participants who had indicated in Q14 that they did not use a corona app, numbers shown in the first data row of Table~\ref{tab:quali-negative}, while Q16 was displayed to all participants.

Across all three countries, the most popular reason not to use a corona app~(Q14a) was that one was (presumed) not to be available (yet) (``no app'', DE: \num{214}, US: \num{222}, CN: \num{73}) or that there were only ones that did not suit participants' particular needs (``no suitable app'', DE: \num{61}, US: \num{18}, CN: \num{13}), such as not having been thoroughly tested or officially recommended by authorities. Another common sentiment was that the app was simply perceived as unnecessary, either in a general way (DE: \num{96}, US: \num{103}, CN: \num{56}) or for more specific reasons, such as the worst phase of the pandemic being over (``state of the pandemic'', DE: \num{8}, US: \num{5}, CN: \num{18}). Unique to China (\num{32}) was the observation that there was no need for a standalone app since coronavirus-related functionality was already offered by plugins for widespread multi-purpose apps like Alipay or WeChat. A similarly distinct reason for Chinese participants not to use a corona app was the fact that the use of such an app was not mandated by the government (``not mandatory'', DE: \num{0}, US: \num{1}, CN: \num{5}), while only participants in Germany and the US questioned the severity of the COVID-19 pandemic in general (``coronaskeptic'', DE: \num{12}, US: \num{16}, CN: \num{0}). Reasons related to personal phone use, such as not \changed{carrying the phone at all times}, were more frequently mentioned in Germany and the US (``phone use'', DE: \num{21}, US: \num{29}, US: \num{1}). The latter could be rooted in the Chinese sample being only representative for the Chinese \emph{online} population.

When we more broadly asked for negative aspects of corona apps (Q16), participants in all countries prominently voiced privacy concerns (DE: \num{292}, US: \num{337}, CN: \num{179}), but German and US participants much more prominently worried about government surveillance (DE: \num{174}, US: \num{70}, CN: \num{1}), their personal autonomy or rights being infringed (DE: \num{36}, US: \num{38}, CN: \num{5}), or discrimination rooted in (non-) use of an app (DE: \num{9}, US: \num{9}, CN: \num{2}). In turn, in the Chinese sample, in which a majority reported to use coronavirus-related functionality on their smartphones, one prominent negative sentiment was \changed{anxiety that the phone could potentially display coronavirus-related warnings at any given time} (DE: \num{12}, US: \num{20}, CN: \num{69}). Still, overall satisfaction with these apps was reported to be high, with \num{232} of Chinese participants stating that there were no negative aspects at all and \num{45} even mentioning positive aspects as an answer to this question.

\subsubsection{Perceived Positive Aspects of Corona Apps}

Our coding frame for perceived positive aspects of corona apps consisted of 20 codes (Krippendorff's alpha: range $(\num{0.45
}, \num{1} )$, weighted mean \num{0.82
}) as depicted in Table~\ref{tab:quali-positive}. The three lowest-performing codes, all in the $(\num{0.45
}, \num{0.47} )$ range, originate from rather unspecific categories (``other'', ``unclear'') and the fact that one code (``symptom check'') was rather rare in the dual-coded segment of the data. 

The answers provide additional hints that people seem to prefer what they are already familiar with, either through public debates or actual use of an app: \num{40} Chinese participants perceived the health certificate functionality to be a positive aspect of corona apps, compared to none in Germany and only one in the US. Another \num{87} responses from China specifically pointed out the use of apps to keep track of people's travel history. While we had not identified this as a separate app purpose in our analysis of existing apps, which included the Chinese health QR code system, the frequency of answers referring to this aspect of the system (rather than the ``health certificate'' part) prompted us to introduce ``movement'' as a distinct code. In Germany and the US, where public discussions of smartphone apps against the pandemic had centered on contact tracing, this purpose was the second most often named positive aspect of corona apps (DE: \num{106}, US: \num{103}) -- directly after ``none'' (DE: \num{212}, US: \num{170}), which was only stated \num{16} times in China. In contrast, almost twice as many Chinese participants (\num{100}) voiced an unspecific but positive 
opinion of coronavirus-related apps as participants from Germany (\num{57}) or the US (\num{56}). This provides another hint that corona apps being in actual use has a positive impact on how people perceive them.

\section{Discussion}

\changed{ 
Our research identified factors that impact people's willingness to adopt mobile apps designed to help fight the {COVID-19} pandemic and provides insights into how these apps are perceived. The results have implications not only for the design of mobile apps developed to help fight the COVID-19 pandemic but also for the design of health-related mobile apps and for apps released in the public interest whose efficiency relies on widespread voluntary adoption.
} 

\changed{
\subsection{Contextual Factors are Important for Privacy Decisions}

The theory of privacy as contextual integrity states that the appropriateness of information flows depends on contextual informational norms governed by actors, transmission principles, and information types. Our study confirms that these factors -- although not all different levels -- play a role in people's decisions to use a corona app. We measured the importance of each factor as the effect on the willingness to use an app and found that different combinations of these factors in fact violate informational norms that differ between countries. For example, sharing any corona-related app data with law enforcement agencies violates informational norms in Germany and the US. Moreover, when the collected data and data sharing are limited and purposes are less invasive, the willingness to use apps increases.
Quite contrary to the idea of CI we also find that the specific information types and transmission principles -- at least in the way they were modeled in our survey -- only play a subordinate role.

Contrary to theories that see informational privacy primarily as a means to limit disclosure~\cite{gurses_surveillance_2009}, our data supports findings from IoT-~\cite{emaminaeini_iot_2017} and health-related privacy research~\cite{baig_dnatesting_2020}: participants are willing to share data when it supports a common good (\ie, ``regional releases of lockdown measures'') over individual benefits, although participants still expect that informational norms are upheld.
}

\changed{
\subsection{People Favor the Familiar, and What is Familiar is Subject to Change}

Our analyses indicate that people favor data processing practices and are more willing to adopt technologies they are already familiar with, such as apps for digital contact tracing in the US and Germany, where the use of smartphone apps for this purpose has been subject of the public debate. In China, a prominently reported positive aspect of corona apps was movement control (Table~\ref{tab:quali-positive}), which had already been put in place there at the time of the survey via the Health Code systems. Chinese participants also favored apps that allowed for unique identification of the individual, reflecting that it is ``standard practice in China''~\cite{voncarnap_china_2020} to send personal data linked to one's national ID to government servers. 
This supports prior research that found people to be more comfortable with allowing data collection in IoT scenarios they perceived to be realistic and ``happen today'' rather than at a later point in the future, and they were more likely to allow data collection for specific ``greater goods'' such as video surveillance for crime prevention~\cite{emaminaeini_iot_2017}. German and US participants' favorable stance towards corona apps that had been positively discussed in the media backs up previous findings that a positive rhetoric around the associated technology can increase people's willingness to share personal data~\cite{gorm_workplace_2016}. This reaffirms that privacy norms evolve~\cite{mulligan_contested_2016} and raises the question how people's stance towards corona apps -- and apps that collect data for a common cause in general -- will change once the {COVID-19} pandemic is over.
}

\subsection{The Future of Apps Against the Spread}

In all surveyed countries, aspects regarding pandemic control were among the most frequently mentioned positive aspects of corona apps. Thus, many citizens seem to be generally open to the idea of using mobile apps to help limiting the spread of a virus.
Privacy concerns were the most frequently mentioned negative aspects of corona apps in all countries. A rigorous approach towards privacy-friendly technologies, which prevent privacy violations a priori, \eg privacy-by-design, seems vital for the success of such apps. 
Somewhat surprisingly, the type of payload data did not have any significant negative influence on the willingness to use a corona app in our study. This implies that users may be willing to share additional data if other conditions are met. Citizens of Germany and the US seem to be willing to use apps for contact tracing without technical malfunctions which ensure anonymity and are administered by the health authorities. Such apps can be further developed to be ready for deployment in potential future pandemics.
In China, apps facilitating individual contact tracing seem to be less relevant. This might be due to the strict regional measures implemented by the authorities as soon as infections are detected and due to the fact that authorities in China already have access to data enabling the tracking of individuals. Instead, Chinese participants expressed an increased need for apps informing about the current (regional or local) regulations and the state of the pandemic as indicated by our results.

\subsection{Implications for the Design of Mobile Health Apps}

Our study was conducted under the impression of a global pandemic and the necessity for apps specifically designed to mitigate a pandemic in the future is unclear.
However, we believe that our results also offer insights for health-related apps outside the specific context of a pandemic.
First, designers and developers of apps that collect sensitive health data should keep in mind the factor levels that positively influence user acceptance (\eg, avoid technical malfunctions, provide data only to health authorities or research institutions, do not collect data that could be used to uniquely identify the user (US and DE)).
Second, many factors (\eg, purpose, identification data, data receiver) showed significant estimates when compared to the baseline, indicating that users would like to be informed about these aspects. They should be clearly communicated to enable informed decisions for or against installation and use of a health-related mobile app.

\subsection{Apps for a Greater Good}

Our findings also provide insights into the question whether people are willing to share personal data with mobile apps developed to serve public interest. In our study the prospect of regional corona restrictions being lifted positively influenced people's willingness to use an app in Germany and the US. 
This was even valued higher than individual freedoms, as represented by the factor level ``the freedom of personal movement or travel,'' and confirms prior findings about people willing to share data if they believe it benefits a good cause~\cite{baig_dnatesting_2020}.

However, our findings also imply that many participants will not support extended use of corona apps beyond the pandemic.
In China, many participants expressed worries that prolonged use of corona apps may increase anxiety due to being in a constant state of alert.
In the US and Germany, where we saw more skepticism towards the governments, almost one third of participants voiced privacy and security concerns (see~Table~\ref{tab:quali-negative}) and many of them were especially worried about surveillance, confirming previous work about apps for {COVID-19} contact tracing~\cite{altmann_acceptability_2020}.
These comments are supported by the negative estimate in our model for scenarios where the data was sent to law enforcement.
These negative sentiments can be alleviated via privacy-preserving approaches in the implementation of ``apps for a greater good'' and transparency about their data processing practices that were found to foster people's trust and increase adoption of {COVID-19} contact tracing apps~\cite{altmann_acceptability_2020, simko_contact_tracing_2020, kaptchuk_covid19apps_2020, zhang_covid_privacy_2020}.
Still, in all countries, we observed a base of participants who reported that they would never use any kind of corona app, regardless of its data processing practices. 
This group may contain people who think that apps are unnecessary to combat the pandemic, fear surveillance, or do not take the threats of coronavirus seriously. 
While previous work~\cite{trang_oneapp_2020} has found a group of critics to be open to arguments especially about privacy, our data shows that a non-negligible number of participants, when prompted to indicate their willingness to use an app, always responded with ``very unlikely''. 
This group is largest in the US ($ n = \num{212} $), followed by Germany (\num{151}), and is rather small in China (\num{8}). 
In the case of contact tracing apps, whose efficiency requires a large adoption rate, this base of 15 to 21\,\% of non-users needs to be taken into consideration when estimating the potential user base of an app, in addition to people who are not able to use any app for other reasons. Similarly, the development process for future apps collecting data to support a public cause needs to take into account that there will always be a base of users unwilling to support any given cause.

\section{Limitations}
In this work we present qualitative and quantitative data allowing detailed insight into the privacy implications and user acceptance of COVID-19-related smartphone apps in three countries. Our approach has the following limitations.

In our survey, we only collect self-reported data, which does not necessarily reflect actual behavior.
This phenomenon (referred to as ``privacy paradox''~\cite{norberg_privacyparadox_2007}) may have been further amplified because all scenarios were hypothetical and participants were asked to imagine themselves in the situation to decide whether or not to use the described app.
By deriving the factor levels used in the scenarios from real-world apps, we aimed to represent the realistic design space for both existing and potential future apps.
Future work may complement ours by conducting field studies that observe participants' actual behavior regarding apps designed to help contain a pandemic.

Four out of five factor levels for the factor ``app purpose'' required specific factor levels for payload data (see Section~\ref{sec:vignette_design}).
This led to a high correlation between app purpose and payload data, which blurs the boundaries between both factors and may have contributed to less relevant and significant results regarding payload data.7

\changed{
For representativity, we focused on age, gender, region, and education.
Ethnicity representation in our US sample is skewed towards the white population (Distribution: \SI{77.7}{\percent} White, \SI{8.9}{\percent} Black or African American, \SI{6.4}{\percent} Hispanic or Latino, \SI{4.3}{\percent} Asian, \SI{2.7}{\percent} Other or no answer).
However, our analysis did not indicate significant effects of ethnicity on the willingness to use a corona app (see Appendix~\ref{appendix:clmm_country_overview}).
}

It is a general problem to reach older, more rural populations in China via online surveys; thus, our panel provider could only provide us with a sample representative for the \emph{online} rather than the general population. 
This leads to the Chinese sample being skewed towards younger age and higher urbanization, which explains the very high rate of smartphone users.

\changed{We commissioned the translation of the Chinese questionnaire and Chinese open-ended answers to a translation agency and followed the quality assurance process of back translation to minimize errors.}
However, we cannot completely exclude the possibility that the translations from and into Chinese may contain small inaccuracies.

Questions asking for already known corona apps (Q13) and individual use of corona apps (Q14) may have been misleading in China as coronavirus-related functionality has been integrated into the established platform economies.
This is indicated by participants responding that they did not use a (dedicated) corona app because there were ``mini programs on WeChat or Alipay'' (participant CN-\num{19}).
While integration into existing platforms can further contribute to app acceptance, we do not expect this potential misunderstanding to have influenced our vignette study as the scenarios for standalone apps were hypothetical.
\section{Conclusion}

The COVID-19 pandemic continues to pose significant challenges to societies around the world. As measures are developed to slow and eventually contain the spread, governments, private companies, and individuals have turned to technology and smartphone apps specifically developed to support these measures. 
To unfold the potential of technological strategies, the population has to support and voluntarily use them instead of rejecting or purposefully work around them because they disagree with the underlying regulations or data collection practices. 
We conducted a study to better understand how privacy factors impact the adoption of coronavirus-related smartphone apps. Based on a vignette design describing various app use cases and scenarios we show that many aspects of the ``contextual privacy'' approach have an actual influence on people's willingness to use a corona app. While in general the willingness to use an already known type of app is high, we found evidence that developers should take additional steps to ensure that apps work correctly, collect as few personal data as possible, and make it transparent when the data will be deleted. Policymakers need to ensure that apps do not share data with private companies, the police, or the public. Our study identified widespread concerns that corona apps could be the start of a new era of government surveillance -- to counter these concerns, strong privacy and security paradigms should be adopted.
We hope that our results can inform the development of novel technology that can help limit the negative impact of COVID-19 and future diseases while respecting users' privacy and autonomy.

\begin{acks}
The authors would like to thank Kangning Zhan for her help with the Chinese translations. 
This research is funded by the Deutsche Forschungsgemeinschaft (DFG, German Research Foundation) under Germany's Excellence Strategy  -- EXC 2092 CASA -- 390781972 and the MKW-NRW Research Training Groups SecHuman and NERD.NRW.

\end{acks}

\bibliographystyle{ACM-Reference-Format}
\bibliography{bibliography/ms}

\appendix

\section{Questionnaire}
\label{appendix:questionnaire}

[Title] Corona App Study 

\subsection*{Welcome Text}
The current situation with the novel coronavirus (SARS-CoV-2) and the disease it causes (COVID-19) has sparked an intense debate about the use of smartphone apps to better understand and contain the spread of the virus.
This study investigates how people perceive apps that promise to help fight the COVID-19 pandemic and what they expect from them.

\subsection*{Phone Use}

First we would like to ask you a few questions about the smartphone you mainly use.

\begin{enumerate}
\item[Q1:] Do you own a smartphone? [single choice]
\begin{itemize}
    \item Yes
    \item No
    \item Prefer not to answer
\end{itemize}

\vspace{0.25cm}
\item[Q2:] What is your phone’s operating system? [single choice]
\begin{itemize}
    \item Android/Google
    \item iOS/Apple
    \item Other (please specify:)
    \item Don’t know
    \item Prefer not to answer
\end{itemize}

\vspace{0.25cm}
\item[Q3:] How satisfied are you with the following aspects of your smartphone? [array of single-choice questions; answer options for each: Very satisfied, Satisfied, Neither satisfied nor dissatisfied, Dissatisfied, Prefer not to answer]
\begin{itemize}
    \item Battery life
    \item Location accuracy (GPS)
    \item Camera quality
    \item Speed (of apps)
\end{itemize}
\end{enumerate}

\subsection*{Coronavirus Experience}

Now we would like to ask you some questions about your experience with the novel coronavirus.\\
This study uses the following terminology:
\begin{itemize}
    \item ``coronavirus'': the novel coronavirus (SARS-CoV-2) that has caused a global pandemic in early 2020;
    \item ``COVID-19'': coronavirus disease 19, the respiratory disease caused by this virus;
    \item ``corona apps'': smartphone apps specifically developed to help limit the spread of the COVID-19 pandemic.
\end{itemize}

\begin{enumerate}

\item[Q4:] Are you or have you been infected with the novel coronavirus? [single choice]
\begin{itemize}
    \item I was tested for coronavirus and at least one of the results was positive.
    \item I was tested for coronavirus and all results were negative.
    \item I was not tested for coronavirus and I do not think I have been infected.
    \item I was not tested for coronavirus, but I suspect that I might have been infected.
    \item Prefer not to answer
\end{itemize}

\vspace{0.25cm}
\item[Q5:] Is there a person in your social circle who is or has been infected with the novel coronavirus? [single choice]
\begin{itemize}
    \item Yes
    \item No
    \item Prefer not to answer
\end{itemize}

\vspace{0.25cm}
\item[Q6:] Have you been quarantined or did you quarantine yourself because of coronavirus? [single choice]
\begin{itemize}
    \item Yes
    \item No
    \item Prefer not to answer
\end{itemize}

\vspace{0.25cm}
\item[Q7:] To the best of your knowledge, is there a person at higher risk in your household, \ie, an older adult or a person of any age who has a serious underlying medical condition? [single choice]
\begin{itemize}
    \item Yes
    \item No
    \item Prefer not to answer
\end{itemize}

\vspace{0.25cm}
\item[Q8:] How concerned are you that you or someone you are close to will become infected with the coronavirus? [single choice]
\begin{itemize}
    \item Not at all concerned 
    \item Slightly concerned 
    \item Somewhat concerned 
    \item Moderately concerned 
    \item Extremely concerned 
    \item Prefer not to answer 
\end{itemize}
\end{enumerate}

\subsection*{Introduction to App Scenarios}

In the following, you will be shown different apps to find out what kind of corona apps you would prefer to use. The presented apps are fictional and have different purposes and implement different functionalities. For each app, you will be asked a few questions. Please consider the app’s purpose and functionalities while answering the questions. Always assume that you are free to choose whether or not you install and use the app.

\subsection*{App Scenario}

[This question group was displayed 10 times, only with different scenarios.]

\paragraph{Sample scenario}

Imagine an app that provides information about your health and needs to be shown if you want to visit a certain place.

\begin{itemize}
    \item The app uses health or activity data, your COVID-19 infection status, and your current or past location(s).
    \item In addition, the app collects data that could be used to uniquely identify you.
    \item This data is sent to research institutions when you request your health report and it will be stored until the current coronavirus regulations end.
    \item The app decreases your phone's battery life. Using this app may increase your personal freedom of movement or travel.
\end{itemize}

\begin{enumerate}

\item[Q9:] How likely are you to use this app on your smartphone? [7-point scale with end points ``Very unlikely'' and ``Very likely'', plus ``Prefer not to answer'']

\vspace{0.25cm}
\item[Q10:] How many people in \{Germany, the United States, China\} do you expect to use this app on their smartphones? [7-point scale with end points ``No one'' and ``Everyone'', plus ``Prefer not to answer'']

\vspace{0.25cm}
\item[Q11:] Please complete the following statement: Most people who are important to me think that I ... [7-point scale with end points ``should not use this app'' and ``should use this app'', plus ``Prefer not to answer'']

\vspace{0.25cm}
\item[Q12:] How useful do you rate this app in helping contain the spread of the COVID-19 pandemic? [7-point scale with end points ``Not at all useful'' and ``Very useful'', plus ``Prefer not to answer'']

\end{enumerate}

\subsection*{Use of Corona Apps}

\begin{enumerate}
\item[Q13:] Do you know of any app recommended by the public authorities in the United States that can be used to ... [array of single-choice questions; answer options for each: Yes, No, Unsure, Prefer not to answer]
\begin{itemize}
    \item ... get information about the novel coronavirus and its spread?
    \item ... check if you have coronavirus-related symptoms?
    \item ... enforce quarantine?
    \item ... identify people you have been in close contact with and alert them in case you tested positive for coronavirus?
    \item ... provide information about your health and needs to be shown if you want to visit a certain place?
\end{itemize}

\vspace{0.25cm}
\item[Q14:] Do you use any kind of corona app on your smartphone? [single choice]
\begin{itemize}
    \item Yes
    \item No
    \item Don’t know 
    \item Prefer not to answer
\end{itemize}

\vspace{0.25cm}
\item[Q14a:] If no: Why do you not use a corona app? [free text]

\vspace{0.25cm}
\item[Q14b:] If yes: Which corona app(s) do you use? [free text]

\vspace{0.25cm}
\item[Q15:] In general, what do you consider positive aspects of smartphone apps to help limit the spread of the COVID-19 pandemic? [free text]

\vspace{0.25cm}
\item[Q16:] In general, what do you consider negative aspects of smartphone apps to help limit the spread of the COVID-19 pandemic? [free text]
\end{enumerate}

\subsection*{Trust in Institutions}

\begin{enumerate}
\item[Q17:] How do you rate the measures taken in your area to fight the COVID-19 pandemic? [single choice]
\begin{itemize}
    \item Way too strict 
    \item Too strict 
    \item About right 
    \item Too lenient
    \item Way too lenient
    \item Prefer not to answer
\end{itemize}

\vspace{0.25cm}
\item[Q18:] What is your overall opinion of the following institutions in the COVID-19 pandemic? [array of single-choice questions; answer options for each: Very favorable, Mostly favorable, Neither favorable nor unfavorable, Mostly unfavorable, Very unfavorable, Prefer not to answer]
\begin{itemize}
    \item Health authorities
    \item Law enforcement
    \item Research institutions
    \item Private companies
    \item Federal government
    \item State government
\end{itemize}

\vspace{0.25cm}
\item[Q19:] In the past private companies have shared their customers’ data, such as phone location data, with public authorities to help limit the spread of the COVID-19 pandemic. How do you rate this practice? [single choice]
\begin{itemize}
    \item Totally unacceptable 
    \item Unacceptable  
    \item Neither unacceptable nor acceptable 
    \item Acceptable 
    \item Perfectly acceptable 
    \item Prefer not to answer 
\end{itemize}

\end{enumerate}

\subsection*{Individual Privacy Concerns}

\begin{enumerate}
\item[Q20:] Please indicate to what extent you agree with each of following statements.

[IUIPC constructs for Control, Awareness (of Privacy Practices), and Collection~\cite{malhotra_iuipc_2004}]
\end{enumerate}




\clearpage
\section{Details of Regression Models}
\label{appendix:clmm_country_overview}

Tables~\ref{tab:clmm_results_vignettes}, \ref{tab:clmm_results}, and \ref{tab:clmm_results2} were computed using the following cumulative link mixed model fitted with the Laplace approximation using following call in R:

\begin{lstlisting}[caption={Function call in R used to fit the cumulative link mixed model. Note that the ``Ethnicity'' variable was only available for the US responses.}]
clmm(Q9 ~ purpose + technical + societal + d1_encounter + d1_location + d1_health + d1_infection + d1_unspec + d1_none + d2 + receiver + retention + transmission + Q2 + Q3r1 + Q3r2 + Q3r3 + Q3r4 + Q4 + Q5 + Q6 + Q7 + Q8 + Q13r1 + Q13r2 + Q13r3 + Q13r4 + Q13r5 + Q14 + Q17 + Q18r1 + Q18r2 + Q18r3 + Q18r4 + Q18r5 + Q18r6 + Q19 + Gender + MainRegionName + Education_General + Age + Ethnicity + iuipc_col + iuipc_awa + iuipc_con + (1 | scenario_index), link = "probit", data = responses, Hess = TRUE)
\end{lstlisting}

Each round $R$ we compared the current best model ($M_R$) with models were each factor (F) was individually removed ($M_{R without F}$) and compared the AIC of these models to decide which factor to remove for the next $M_R$
Factors were removed in the following order:

\begin{lstlisting}[caption={Factors removed of the final model from US responses in order of removal}]
Ethnicity, Q18r4, Q17, Q13r3, Q3r4, Q3r3, Q13r1, d1_location, d1_unspec, d1_none, Q18r5, Q7, Q6, d1_encounter, Q2, iuipc_awa, d1_health, iuipc_col, Education_General, d1_infection
\end{lstlisting}

\begin{lstlisting}[caption={Factors removed of the final model from Chinese responses in order of removal}]
Q3r4, Age, societal, Q18r6, Q18r5, Q13r5, Q17, Q13r1, Q4, Q3r2, d1_none, Q2, Gender, Q5, MainRegionName, Q3r3, d1_encounter, transmission, d1_infection
\end{lstlisting}

\begin{lstlisting}[caption={Factors removed of the final model from German responses in order of removal}]
MainRegionName, Q3r3, Age, Q13r1, Q3r1, Q6, Q3r4, d1_none, iuipc_con, Q7, retention, d1_unspec, Q18r2, Q4, Age, Q5, d1_health, iuipc_awa, Education_General
\end{lstlisting}

\end{document}